\documentclass{article}

\usepackage{arxiv}

\usepackage[svgnames]{xcolor}

\usepackage[utf8]{inputenc}
\usepackage{amsmath}             
\usepackage{dsfont}

\usepackage{pgfplots}
\pgfplotsset{
 compat=1.11,
}

\usepackage{float}
\usepackage{multirow}
\usepackage{graphicx}
\usepackage{caption}
\usepackage{subcaption}

\usepackage[numbers]{natbib}
\usepackage{amssymb}
\usepackage[colorlinks,allcolors=black,pdftex]{hyperref}

\usepackage[nottoc]{tocbibind}
\usepackage{tikz-qtree}

\usepackage[export]{adjustbox}

\title{Tree models for assessing covariate-dependent method agreement}
\author{Siranush Karapetyan \\
Institute of General Practice and Health Services Research \\
Technical University of Munich \\
\texttt{Siranush.Karapetyan@mri.tum.de} \\
\And
Achim Zeileis \\
Faculty of Economics and Statistics \\
University of Innsbruck \\
\texttt{Achim.Zeileis@uibk.ac.at} \\
\And
André Henriksen \\
Department of Computer Science \\
The Arctic University of Norway \\
\texttt{Andre.Henriksen@uit.no} \\
\And
Alexander Hapfelmeier \\
Institute of General Practice and Health Services Research \\
Institute of AI and Informatics in Medicine \\
Technical University of Munich \\
\texttt{Alexander.Hapfelmeier@mri.tum.de} \\
}

\begin{document}

\maketitle

\begin{abstract}
Method comparison studies explore the agreement of measurements made by two or more methods. Commonly, agreement is evaluated by the well-established Bland-Altman analysis. However, the underlying assumption is that differences between measurements are identically distributed for all observational units and in all application settings. We introduce the concept of conditional method agreement and propose a respective modeling approach to alleviate this constraint. Therefore, the Bland-Altman analysis is embedded in the framework of recursive partitioning to explicitly define subgroups with heterogeneous agreement in dependence of covariates in an exploratory analysis. Three different modeling approaches, conditional inference trees with an appropriate transformation of the modeled differences (CTreeTrafo), distributional regression trees (DistTree), and model-based trees (MOB) are considered. The performance of these models is evaluated in terms of type-I error probability and power in several simulation studies. Further, the adjusted rand index (ARI) is used to quantify the models' ability to uncover given subgroups. An application example to real data of accelerometer device measurements is used to demonstrate the applicability. Additionally, a two-sample Bland-Altman test is proposed for exploratory or confirmatory hypothesis testing of differences in agreement between subgroups. Results indicate that all models were able to detect given subgroups with high accuracy as the sample size increased. Relevant covariates that may affect agreement could be detected in the application to accelerometer data. We conclude that conditional method agreement trees (COAT) enable the exploratory analysis of method agreement in dependence of covariates and the respective exploratory or confirmatory hypothesis testing of group differences. It is made publicly available through the R~package \texttt{coat}.
\end{abstract}


\section{Introduction}
\label{sec:intro}
Method comparison studies are relevant in all scientific fields whenever the agreement of continuously scaled measurements made by two or more methods is to be investigated. However, they have found particular application in medical research, for example in laboratory research \citep[]{Chhapola2015, Giavarina2015}, anaesthesiology \citep[]{Abu2016}, ophthalmology \citep[]{Bunce2009} and pathology \citep[]{Jensen2006} among many others. Here, taking measurements can be time-consuming, expensive, invasive or stressful for patients. Therefore, methods are constantly being developed and improved to reduce these shortcomings \citep[][]{Bunce2009}. However, the agreement between a new method and a standard method needs to be shown in order to replace  the latter. A well-established methodology for analysis was developed by Bland and Altman and is known as the Bland-Altman analysis or plot \citep[]{1_altman1983measurement}. In its most basic form, it illustrates the differences against the mean values of paired measurements made by two methods. Here, two quantities of interest are the mean difference, referred to as `bias', and the standard deviation of the differences, which is used to determine the width of the so called `Limits of Agreement' (LoA) \citep[]{Hanneman2008}. The bias is a measure of the overall deviation of the methods but has limited interpretability, since large positive and negative deviations can still add up to a small overall bias. Further, as a summary measure, the expected agreement of a single subject's measurements cannot be inferred from the bias. Therefore, Bland and Altman proposed to estimate the LoA, that is a prediction interval in which about $95\%$ of individual differences between the measurements of the two methods are expected to lie. The mean and standard deviation of differences can be calculated directly from the observed data but it has also been suggested to use regression modeling under the assumption of normally distributed residuals \citep[]{Carstensen2010comparing, 6_carstensen2011comparing}.

Proper planning, conduct, interpretation and reporting of method comparison studies has been the subject of ongoing research and recommendations have been provided in respective publications and through reviews of the relevant literature \citep[]{Stoeckl2004, Jensen2006, Bunce2009, Hanneman2008, Giavarina2015, Chhapola2015, Abu2016, hapfelmeier2016cardiac, Gerke2020}. These works are also concerned with the data description, processing and analysis, the plotting of results, the (pre)specification of acceptable agreement, the precision of estimation, the repeatability of measurements and the investigation of homoscedastic variances and trends. Regarding the latter two, Bland and Altman already discussed early the question whether the agreement between the methods depends on the magnitude of the measured values, that is whether there is a relationship between the differences and the means of paired values \citep[]{1_altman1983measurement, 2_bland1986statistical}. In that case, they suggested either transforming (e.g. log-transforming) the differences to remove the dependency or modeling the differences with mean values as explanatory variable in a linear regression model.

In the present work, we suppose that the underlying assumption of a Bland-Altman analysis, that is that the agreement of methods is identically distributed for all observational units or subjects, may not be valid in any case. The basic idea is that the methods' measurements can be affected by internal and external factors, such as the subjects' characteristics and measurement settings, with direct implications on the agreement of methods. Previous studies have used heuristic approaches to address this issue, for example through the post-hoc fitting of additional regression models and subgroup analyses \citep[]{9_huber2014room, haghayegh2020comprehensive}. An early example is the regression of mean values on differences as originally suggested by Bland and Altman and outlined above \citep[]{1_altman1983measurement, 2_bland1986statistical}.

Here, we introduce a unifying framework and analysis approach for conditional method agreement in case of single measurements per subject or observational unit. Recursive partitioning is used to simultaneously explore relations between covariates and agreement and to define corresponding subgroups with heterogeneous agreement in terms of bias and/or the width of LoA, taking advantage of the fact that a Bland-Altman analysis can be parameterized accordingly \citep[]{Carstensen2010comparing, 6_carstensen2011comparing, Moller2021}. We consider three different modeling approaches, that is conditional inference trees with an appropriate transformation of the outcome (CTreeTrafo) \citep[]{10_hothorn2006unbiased}, distributional regression trees (DistTree) \citep[]{12_schlosser2019distributional}, and model-based trees (MOB) \citep[]{11_zeileis2008model}. The ability of these approaches to control the type-I error probability at a nominal level, the power to detect given subgroups, and the ability to accurately define these subgroups is investigated in simulation studies. We also demonstrate the relevance to medical research through applications to a real data example of accelerometer measurements made by different devices. In addition, we propose a two-sample Bland-Altman test suitable for exploratory or confirmatory hypothesis testing of differences in agreement between two (pre)defined subgroups.

\section{Methods} \label{Meth}
The following subsections outline the concept of conditional method agreement, corresponding modeling through recursive partitioning and a two-sample Bland-Altman test for hypothesis testing of group differences in agreement. The models used for analysis are called conditional method agreement trees (COAT).

\subsection{Introductory example}
\begin{figure}
  \centering
\begin{subfigure}[b]{0.4\textwidth}
    \includegraphics[width=\textwidth,valign=c]{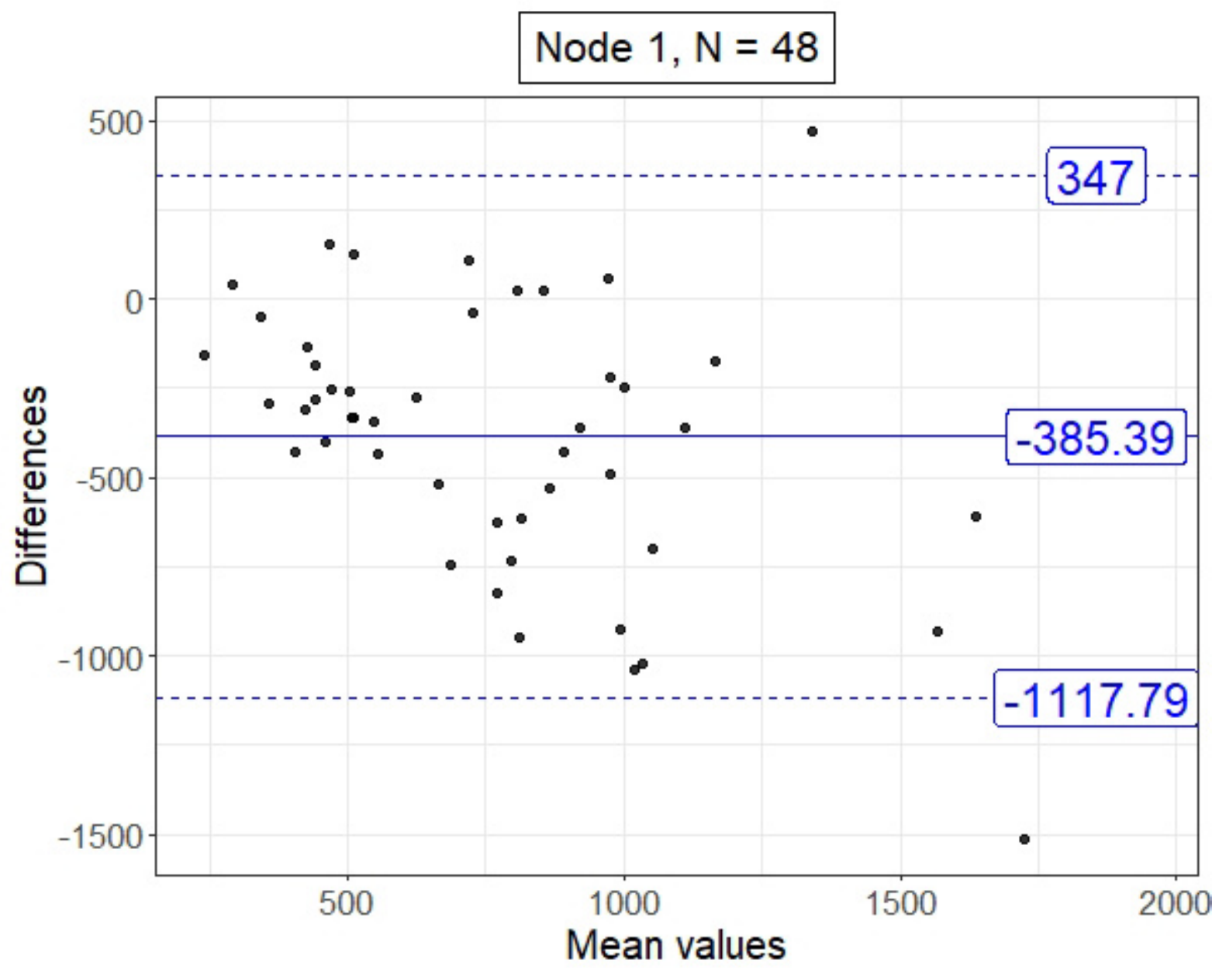}
    \subcaption{Bland-Altman plot}
    \label{tab:COATDEV_A}
\end{subfigure}
\begin{subfigure}[b]{0.4\textwidth}
   \includegraphics[width=1.1\textwidth,valign=c]{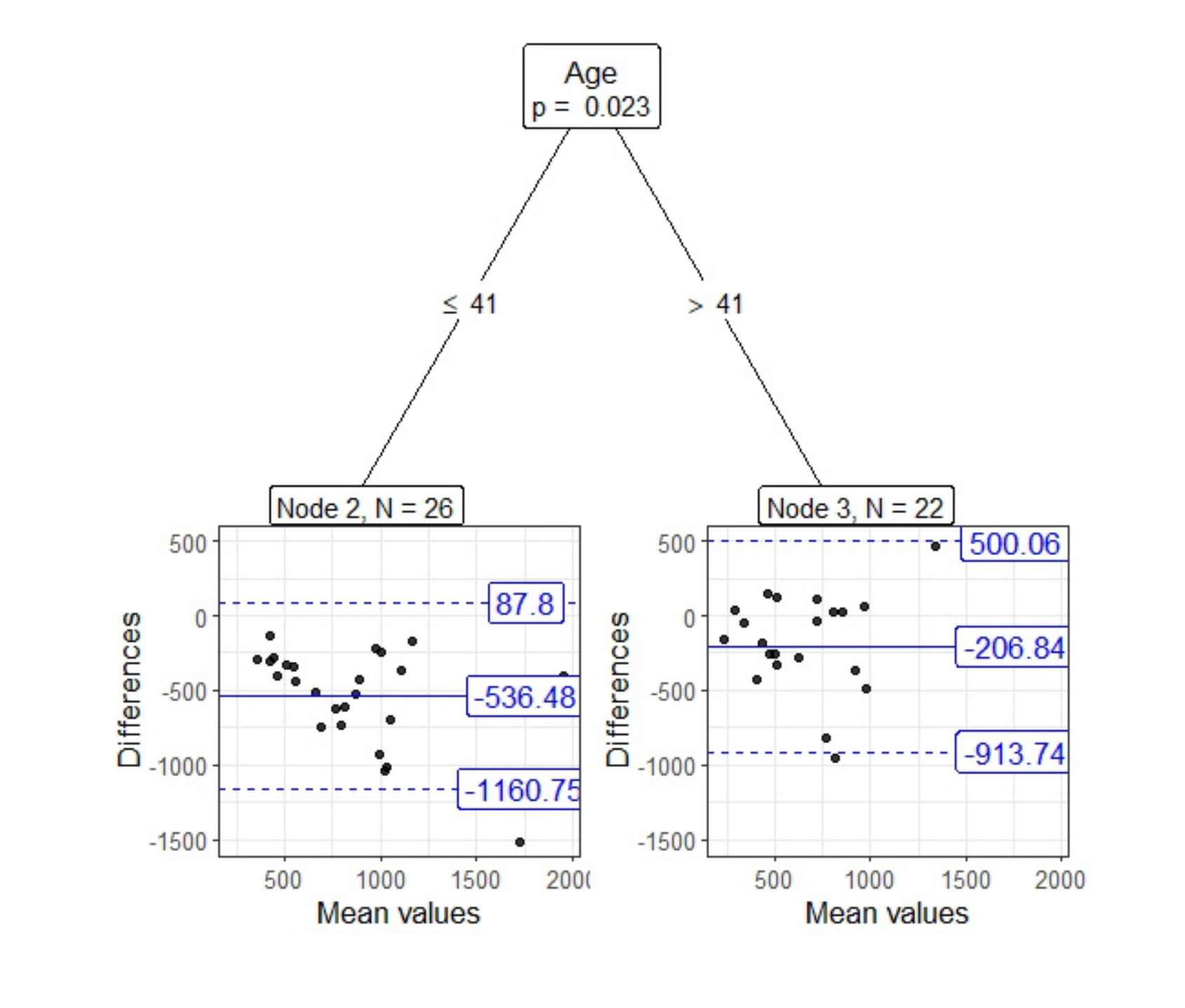}
   \subcaption{COAT plot}
   \label{tab:COATDEV_B}
\end{subfigure}
  \caption{Agreement (\ref{tab:COATDEV_A}) and conditional agreement (\ref{tab:COATDEV_B}) of activity energy expenditure (AEE) (kilocalories) measured by two different accelerometers.}
  \label{tab:COATDEV}
\end{figure}

The concept of conditional method agreement is briefly illustrated here using a real data example, which is explained in more detail in Section~\ref{sec:data}. The data consists of 24-hour accelerometer measurements and socio-demographic information from $n=50$ participants of the original study \citep[]{henriksen2019validity}. Figure~\ref{tab:COATDEV_A} shows a respective Bland-Altman plot of the agreement of activity energy expenditure (kilocalories) measured by two investigated devices. Using COAT by MOB, it can be shown that this agreement is related to the age of the participants. There are two subgroups with statistically significantly different agreement ($p=0.023$), especially in terms of bias, which is divided from $-385$ in the whole sample into $-536$ and $-207$ in the subgroups defined by a split point of $41$ years (cf. Figure~\ref{tab:COATDEV_B}). Also, the LoA within the defined subgroups are less wide than for the whole sample. Comparing the subgroups, the LoA are wider within subjects of increased age of $>41$ years. This result can be of interest to scientists, health professionals, users and manufacturers of accelerometers who develop the devices or rely on their functionality and who may want to discuss the reasons for this difference in agreement and possible solutions or implications for proper use.

\subsection{Conditional method agreement}
\label{sec:coat}
As shown in the previous example and discussed in Section~\ref{sec:intro}, in a Bland-Altman analysis we are essentially interested in the first and second moments of the marginal density function $f_{Y}(y)$. Here, $Y=M_{1}-M_{2}$ is a random variable of independent and identically distributed (iid) differences between two methods' paired measurements $(M_{1}, M_{2})$. The moments of $f_{Y}(y)$ are the expectation $\mathbb{E}(Y)$ and the variance $\mathrm{Var(Y)} $ with corresponding estimates given by the mean $\Bar{y} = \sum_{i=1}^{n}y_{i}$ and the empirical variance $s^{2} = \frac{1}{n-1}\sum_{i=1}^{n} (y_{i}-\Bar{y})^{2}$ of the observed differences $y_{i}$, $i \in \{1, \ldots, n\}$, of $n$ subjects or observational units. Thereby, $\Bar{y}$ describes the overall deviation between methods, which is often referred to as the `bias' \citep[]{Hanneman2008}. However, as discussed in Section~\ref{sec:intro}, the bias is of limited use because it does not provide information about the individual agreement of measurements made for the same subject or observational unit. Interpretation of agreement in a Bland-Altman plot therefore relies mainly on 95\% prediction intervals, that is the LoA, which are calculated using $\Bar{y}$ and $s$ with an appropriate distributional assumption about $Y$. Most often, a normal distribution or t distribution is assumed.

Another assumption of a Bland-Altman analysis is that $\mathbb{E}(Y)$ and $\mathrm{Var(Y)}$ are independent of the magnitude of measurements, implying that the differences are iid. However, if the observed distribution of data suggests that such an association has to be assumed, Bland and Altman propose either to remove this relationship by transforming the differences, for example to establish homoscedasticity by using a log-transformation, or to use a regression model considering the differences $Y$ as the outcome and the mean measurements $M=\frac{1}{2}(M_{1}+M_{2})$ as an explanatory variable \cite{3_bland1999measuring}. We generalize this approach to define conditional method agreement as follows. 

Given a random variable $Y=M_{1}-M_{2}$ of differences between two methods' measurements $M_{1}$ and $M_{2}$ and a multivariable vector of covariates $X$, conditional method agreement can be formalized as
$$
f_{Y}(y|x) \neq f_{Y}(y).
$$

Here $f_Y(y|x)$ is the conditional density function of $Y$ given $X=x$. The realizations $y$ are the observed differences and $x$ are the measured covariate values which can also include mean values $m=\frac{1}{2}(m_{1}+m_{2})$ of paired measurements. In the present work, we use COAT to obtain estimates of the conditional expectation $\mathbb{E}(Y|X)$ and the conditional variance $\mathrm{Var(Y|X)}$ to assess conditional method agreement, with and without using distributional assumptions about $f_{Y}(y|x)$. Respective null-hypotheses
\begin{equation} \label{eq:H0_both}
H_{0}: \mathbb{E}(Y|X) = \mathbb{E}(Y) \ \cap \ \mathrm{Var(Y|X)} = \mathrm{Var(Y)}
\end{equation}
or each of
\begin{equation} \label{eq:H0_each}
H_{0}: \mathbb{E}(Y|X) = \mathbb{E}(Y) \ \ \text{and} \ \ H_{0}: \mathrm{Var(Y|X)} = \mathrm{Var(Y)}
\end{equation}

are tested by COAT to determine the statistical significance of conditional estimates. The procedure can also be used to perform a two-sample `Bland-Altman test' to compare agreement between subgroups.

\subsection{Recursive partitioning of method agreement}

The general idea of recursive partitioning is to assess sequentially whether an investigated outcome variable (or model) is homogeneous across all available covariates and, if this is not the case, to capture the differences by splits into more homogeneous subsets of the data \citep[]{breiman2017classification}. The procedure continues recursively until some kind of stopping criterion is reached. The resulting model is often referred to as a tree because of its structure. The subsets considered for splitting or emerging from splitting are termed parent nodes or daughter/child nodes, respectively. A so called stump is obtained if a single split is performed. The definition of the splits performed in the covariates provides decision rules that specify the subsets. 

To define heterogeneous subsets in terms of $\mathbb{E}(Y|X)$ and $\mathrm{Var(Y|X)}$, referring to the mean (bias) and standard deviation of the differences $y$, we consider the following tree-based algorithms: conditional inference tree with an appropriate transformation of the outcome (CTreeTrafo) \citep[]{10_hothorn2006unbiased}, distributional tree (DistTree) \citep[]{12_schlosser2019distributional}, and model-based recursive partitioning (MOB) \citep[]{11_zeileis2008model}. All of these modeling approaches are based on the same basic steps \citep[]{13_schlosser2019power}:

\begin{enumerate}
    \item A model is fit to the entire data by optimizing some objective function or a transformation function is defined.
    \item A split variable is selected based on the association of some goodness-of-fit measure with each possible variable. The variable with the highest significant association is selected.
    \item A split point is chosen so the goodness-of-fit is maximized in the resulting subsets.
    \item Steps $1.-3.$ are repeated until no more significant associations are found or the resulting sample is too small for further splits.
\end{enumerate}

The basic algorithm of the three models considered is thus similar. However, they differ in the implementation of the individual steps, as explained in more detail in the following. Default features of all of the aforementioned models are summarized in Table \ref{tab:Model_Differences}.

\begin{table}
   \centering
   \begin{tabular}{lllll}
         \hline
         & Fit & Test & Statistic & Transformation \\ \hline
         CTreeTrafo & non-parametric & permutation & quadratic & $(y_i, (y_i - \overline{y}_{\omega})^2)$ \\
         DistTree & parametric & permutation & quadratic & $s(\boldsymbol{\hat{\theta}}, y_i)$ \\
         MOB & parametric & fluctuation & quadratic & $s(\boldsymbol{\hat{\theta}}, y_i)$ \\
        \hline
    \end{tabular}
   \caption{Characteristics of the considered COAT models.}
  \label{tab:Model_Differences}
\end{table}

\subsubsection{Conditional inference tree}
\label{sec:ctree}
The algorithm uses permutation tests \citep[]{strasser1999asymptotic, 10_hothorn2006unbiased}, asymptotic by default, to explore whether there is a statistically significant dependence of the outcome on a covariate. Therefore, $j$ partial hypotheses of independence $H_0^j: f_{Y}(y|x_j) = f_{Y}(y)$ are defined for $j=1, ..., J$ covariates. The respective linear test statistics is

\[
t_j = vec \left( \sum_{i=1}^n \omega_i g_j(x_{ji}) h(y_i, (y_1, ..., y_n))^\top  \right)  \in \mathds{R}^{pq},
\]

where $\omega_i$ is a case weight of zero or one, indicating the correspondence of an observation to the node or subset in which the test is performed. $g_j(\cdot)$ and $h(\cdot)$ represent non-random transformation functions. The choice of $g_j(\cdot)$ depends on the type of the $j$-th covariate. The identity function, $g_j(x_{ji}) = x_{ji}$, is a natural choice for a continuous variable, while the indicator function $g_j(x_{ji}) = (I(x_{ji}=1), ..., I(x_{ji}=K))$ is more appropriate for a categorical variable with $K$ levels. With the $vec(\cdot)$ operator, the test statistic becomes a $pq$ column vector, where $p=K$ for categorical covariates and $p=1$ for continuous covariates with identity transformation. $q$ depends on the choice of $h(\cdot)$ and takes a value of $2$ in our case, as outlined below.

In the present setting, that is to model method agreement through the estimation of $\mathbb{E}(Y|X)$ and $\mathrm{Var(Y|X)}$, we define $h(\cdot) = (y_i, (y_i - \overline{y}_{\omega})^2)$, which corresponds to the first step in the basic algorithm. The respective test statistic $t_j$ is then defined as 

\[
t_j = vec \left( \sum\limits_{i=1}^n \omega_i g_j(x_{ji}) (y_i, (y_i - \overline{y}_{\omega})^2)^\top  \right) \in \mathds{R}^{p2},
\]

where $\overline{y}_{\omega} = \sum_{i=1}^n \omega_i y_i / \sum_{i=1}^n \omega_i$ is the mean outcome in the node or subset in which the test is performed. The conditional expectation $\mu_j$ and the covariance $\Sigma_j$ of $t_j$ under the null hypothesis $H_0^j$ can be used to obtain the standardized test statistic 

\[
c_{max}(t_j, \mu_j, \Sigma_j) = \max_{z=1,...,p2} \left | \frac{(t_j-\mu_j)_z}{\sqrt{(\Sigma_j)_{zz}}} \right |,
\]

which follows an asymptotic normal distribution. As an alternative, a quadratic form 

\[
c_{quad}(t_j, \mu_j, \Sigma_j) = (t_j-\mu_j) \Sigma_j^+ (t_j-\mu_j)^\top,
\]

can also be used, where the asymptotic conditional distribution is $\chi^2$ with degrees of freedom given by the rank of $\Sigma_{j}$. $\Sigma_j^+$ is the Moore-Penrose inverse of $\Sigma_j$. Standardization of the linear test statistic enables the computation of a $p$-value, where $H_j^0$ can be rejected if this value falls below a specified significance level. The $j$-th covariate with the minimum and statistically significant $p$-value is selected for splitting, corresponding to the second step in the basic algorithm. Note, that the multiple testing problem is present, as hypotheses for several covariates are checked. Therefore, the CTree algorithm uses Bonferroni-adjusted $p$-values by default \citep[]{10_hothorn2006unbiased}.

After selecting the split variable $j^{*}$, the subsequent and third step of the basic algorithm is to find the optimal split point in a continuous variable or dichotomization of the $K$ categories of a categorical variable for binary splitting, which is again determined through a linear test statistic

\[
t_{j^*}^A = vec \left( \sum\limits_{i=1}^n \omega_i I(x_{j^{*}i \in A}) (y_i, (y_i - \overline{y}_{\omega})^2)^\top  \right)  \in \mathds{R}^{2}.
\]

Here, $t_{j^*}^A$ implicitly measures the discrepancy between the subsets $\{y_i|\omega_i = 1$ and $x_{j^{*}i} \in A; i=1,...,n\}$ and $\{y_i|\omega_i = 1$ and $x_{j^{*}i} \notin A; i=1,...,n\}$ in terms of a metric defined by $h(\cdot)$. The best split point is found by maximizing

$$
A^* = \arg \max_{A} c(t_{j^*}^A, \mu_{j^*}^A, \Sigma_{j^*}^A),
$$

over all possible subsets $A$ using the conditional expectation $\mu_{j^*}^A$ and covariance $\Sigma_{j^*}^A$ of $t_{j^*}^A$. This procedure is recursively repeated until no further statistically significant associations are found or subsets become too small for further splitting (which is the fourth step in the basic algorithm).

\subsubsection{Distributional tree}
\label{sec:distttree}
DistTree is similar to CTreeTrafo while a parametric model is fit to the data and the transformation function is replaced with the resulting score function. In particular, DistTree models all parameters of a given distribution \citep[][]{12_schlosser2019distributional}. In the present setting, it is reasonable to assume a normal distribution with the location and scale parameters $\mu$ and $\sigma^{2}$ for the differences $Y$ \citep[]{2_bland1986statistical}. This allows the specification of the corresponding log-likelihood

$$
l(\boldsymbol{\theta}; Y) = \log \left \{ \frac{1}{\sigma \sqrt{2\pi}} \phi \left(\frac{Y - \mu}{\sigma}\right) \right \}; \quad \boldsymbol{\theta} = (\mu, \sigma),
$$

and its score function $s(\boldsymbol{\theta}, Y) = \partial l(\boldsymbol{\theta}; Y) / \partial \boldsymbol{\theta}$ as a measure of goodness-of-fit. $\phi(\cdot)$ is the density function of a standard normal distribution. A maximum likelihood (ML) estimate of $\boldsymbol{\theta}$ is $\boldsymbol{\hat{\theta}} = \arg\max \sum_{i=1}^n l(\boldsymbol{\theta}; y_i).$ This corresponds to the first step in the basic algorithm.

When it is assumed that the differences y are not iid, DistTree can be used to model the conditional expectation $\mathbb{E}(Y|X)$ and variance $\mathrm{Var(Y|X)}$. To do so, a possible association of $\boldsymbol{\theta}$ and a covariate $X_j$ is tested in terms of the null-hypothesis $H_0^j: s(\boldsymbol{\theta}, Y) \perp X_j$, based on the test statistic

\[
t_j = vec \left( \sum\limits_{i=1}^n g_j(x_{ji}) s(\boldsymbol{\hat{\theta}}, y_i)  \right).    
\]

Here, $\boldsymbol{\hat{\theta}}$ is substituted into the score function to obtain $s(\boldsymbol{\hat{\theta}}, y_{i})$ as a measure of goodness-of-fit for each of the observations $y_{i}$. The transformation function $g_j$, as well as the standardized test statistics $c_{quad}(t_j, \mu_j, \Sigma_j)$ and $c_{max}(t_j, \mu_j, \Sigma_j)$ are defined as outlined in Section~\ref{sec:ctree}. The split variable $X_{j^*}$ is determined by the lowest and statistically significant p-value, which is by default corrected for multiple testing (equals step 2 of the basic algorithm). In the third step the split point is chosen so that it leads to the largest discrepancy in the sum of scores between the resulting subsets. This procedure is repeated recursively in each subset until no further significant associations are found or the resulting subsets become too small for further splitting.

It is important at this point to draw attention to the similarity of the statistics $t_{j}$ of CTreeTrafo and DistTree, with CTreeTrafo using a transformation function $h(\cdot)$ instead of the score function $s(\cdot)$ in the calculation. We show the equality of the resulting quadratic test statistics $c_{quad}(\cdot)$ of CTreeTrafo (with the transformation function $h(\cdot)$ defined as given in the previous Section~\ref{sec:ctree}) and DistTree analytically for the case of a continuous predictor in appendix \ref{TestStat}.

\subsubsection{Model-based recursive partitioning}
MOB is similar to DistTree, but uses a different underlying model and hypothesis test. MOB uses fluctuation tests for parameter instability in regression model fits to build a tree model \citep[]{zeileis2007generalized}. In the first step of MOB, a parametric model is fit to the data by maximum likelihood estimation. In the present case, we consider an intercept-only linear regression model $y_{i} = \beta_{0} + \epsilon_{i}$, $\epsilon_{i} \sim \mathcal{N}(0, \sigma)$, to obtain estimates of the expectation $\mathbb{E}(Y) = \beta_{0}$ and variance $\mathrm{Var(Y)} = \sigma^2$. The second step is to assess parameter instability of the estimated model parameters $\hat{\theta} = (\widehat{\beta}_{0}, \widehat{\sigma})$ across the values $x_{j}$ of a potential split variable $X_{j}$. Instability is concluded when the scores $s(\boldsymbol{\hat{\theta}}, y_i)$ do not fluctuate randomly along the ordered values $x_{j}$ \citep[see][for details]{11_zeileis2008model}. The split variable $X_j^*$ is selected as it provides the minimal and statistically significant $p$-value, which is by default corrected for multiple testing. The split in $x_j^*$ is determined so it maximizes the sum of the log-likelihoods of models that are refit to the resulting subsets, corresponding to the third step in the basic algorithm. As with CTreeTrafo and DistTree, the procedure is repeated recursively in each subset until no further significant associations are found or the resulting subsets become too small for further splitting.

\section{Simulation studies}
\subsection{Design}
We conduct simulation studies to investigate the performance of COAT. For each of the defined scenarios, we run $10000$ simulations, and consider sample sizes $n \in \{50, 100, 150, \ldots, 1000\}$, which are common in medical research. A CTree with the default transformation function $h(y_{i}, (y_{i}, \ldots, y_{n})) = y_{i}$, as implemented through the function \texttt{ctree()} of the \texttt{R} package \texttt{partykit} \citep[]{hothorn2015partykit}, is used as a benchmark. Due to the equivalence of the statistics $t_j$ of CTreeTrafo and DistTree, they are also referred to jointly as CTreeTrafo/DistTree in the following.

The assessment of performance is based on the type-I error and the power to reject $H_{0}$ as defined in (\ref{eq:H0_both}) and (\ref{eq:H0_each}), and the Adjusted Rand Index (ARI). The latter is a measure of concordance of two classifications \citep[]{hubert1985comparing}, as it quantifies the proportion of paired observations that belong to the same or different class levels in either classification among the total number of paired observations \citep[]{rand1971objective}. In the case of independent or random classifications, the ARI takes a value of $0$. Higher values indicate a higher concordance, with $1$ indicating perfect agreement. In the present simulations studies, the ARI is used to assess the concordance between the given subgroups and the subgroups defined by COAT. Three different simulation scenarios are considered as follows.

In the Null Case, the method agreement does not depend on any covariates. The simulated data consists of six independent, standard-normally distributed variables including the outcome $Y$, which is the simulated differences between the methods, and five uninformative covariates $X$. The Null Case allows the exploration of the type I error as we look for statistically significant p-values in the root nodes of the COAT models that were fit to the simulated data. The nominal significance level is set to $\alpha=0.05$.

The Stump Case covers three different scenarios. In each of them there are five standard-normally distributed covariates $X$, where method agreement depends on the informative covariate $X_{1}$ such that $Y \sim \mathcal{N}(\mu_{k}, \sigma_{k})$, $k \in \{1,2,3\}$, where

$$
(\mu_{k}, \sigma_{k}) = \begin{cases}
			(\mu_{1} = 0.3 \cdot I(X_{1} > Q_{0.25}), \sigma_{1} = 1) & \text{if $k=1$,}\\
            (\mu_{2} = 0, \sigma_{2} = 1+I(X_{1} > Q_{0.25})) & \text{if $k=2$,}\\
            (\mu_{3} = 0.4 \cdot I(X_{1} > Q_{0.25}), \sigma_{3} = 1+I(X_{1} > Q_{0.25})) & \text{if $k=3$.}
		 \end{cases}
$$

\begin{figure}
    \centering
\begin{subfigure}[b]{1\textwidth}
  \includegraphics[width=1.1\textwidth,valign=c]{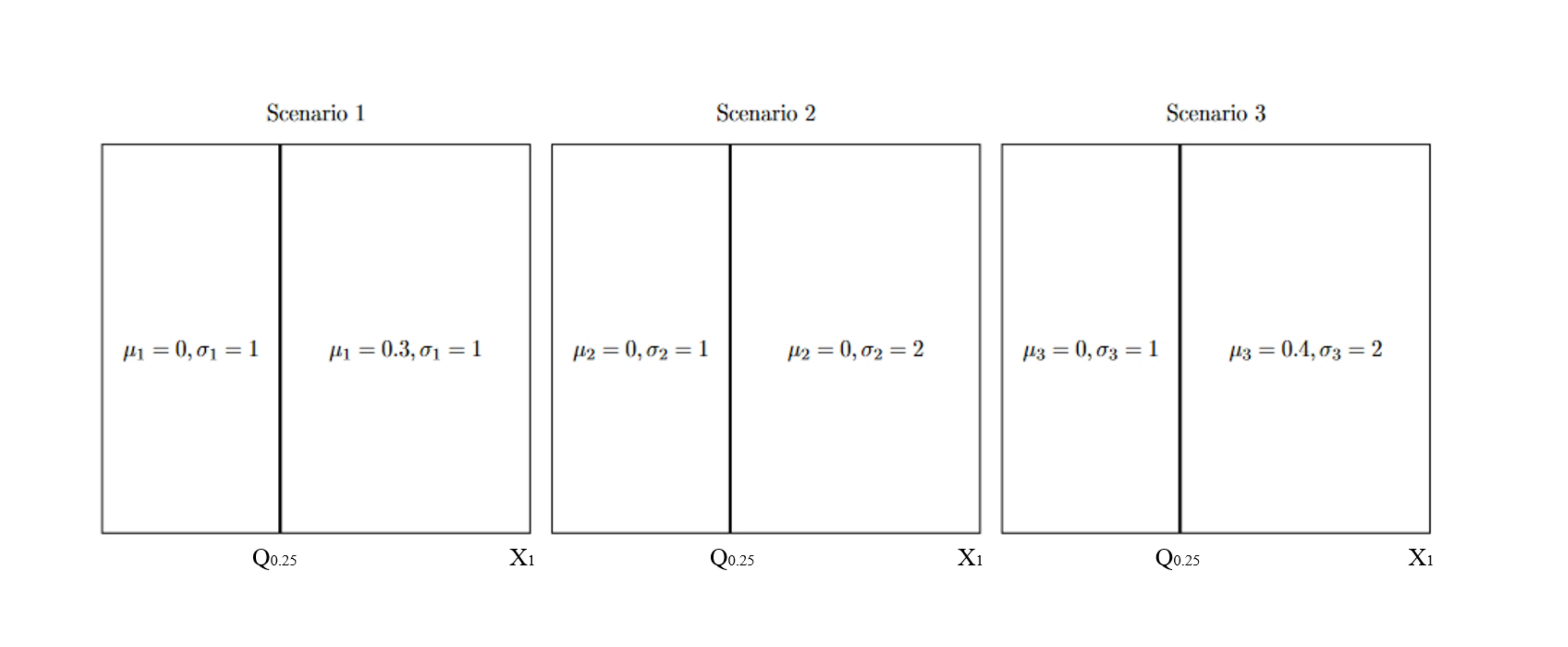}
 \subcaption{Illustration of the Stump Case}
 \label{rec:stump}
\end{subfigure}
\begin{subfigure}[b]{0.8\textwidth}
  \includegraphics[width=1.1\textwidth,valign=c]{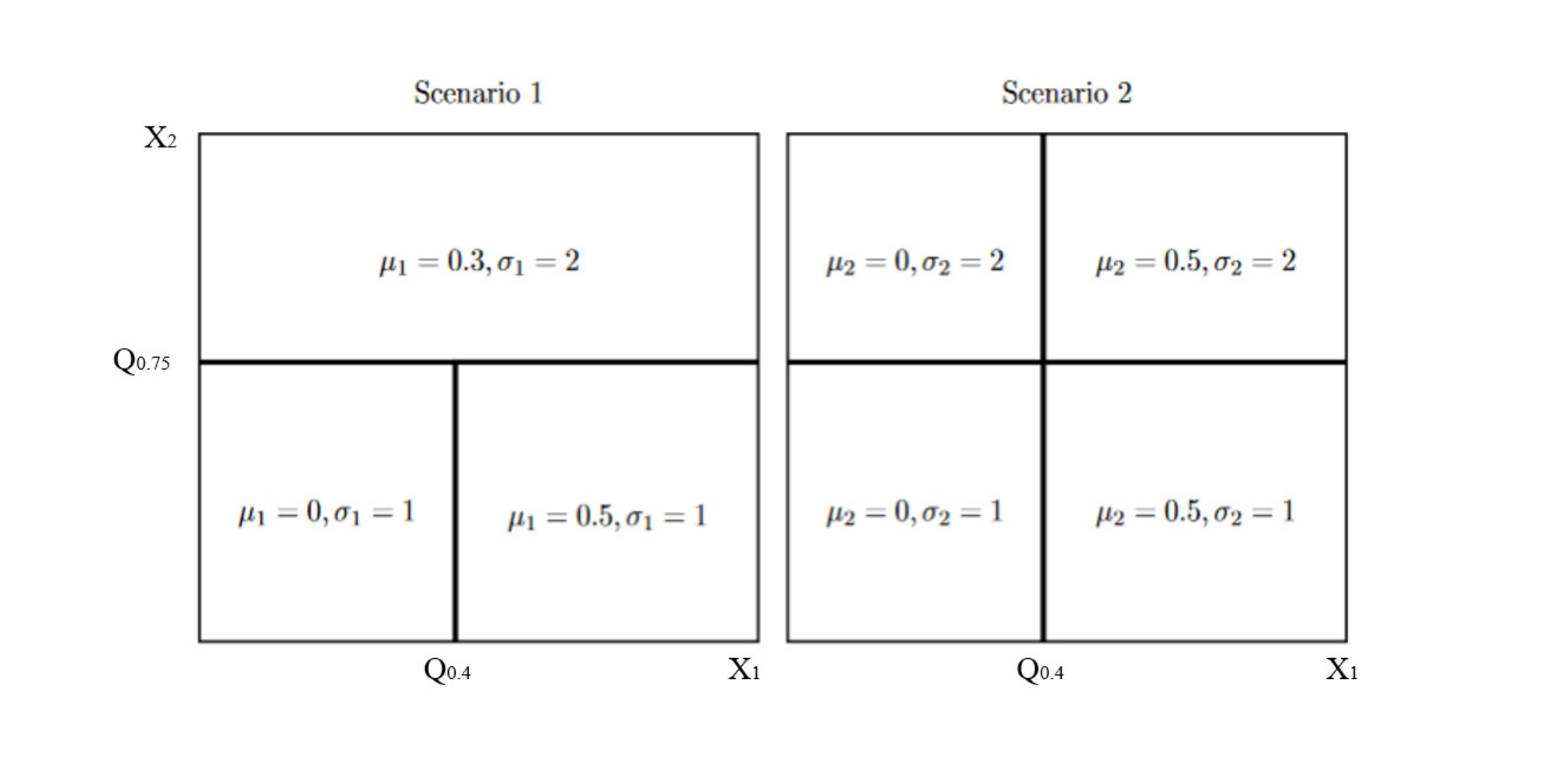}
 \subcaption{Illustration of the Tree Case}
 \label{rec:tree}
\end{subfigure}
\caption{Partitions of $X$ used to define the subgroups in the simulation studies.}
\label{fig:sim}
\end{figure}

Here, $Q_{0.25}$ is the $25$th percentile of the standard normal distribution and has been chosen as a split point in $X_{1}$ to create subgroups that approximately comprise $25\%$ and $75\%$ of the observations. The subgroups consequently differ only in $\mu_{k} = \mathbb{E}(Y|X)$, that is in the bias of method agreement in the scenario $k=1$, they differ in $\sigma_{k} = \mathrm{Var(Y|X)}$, that is in the width of the LoA in the scenario $k=2$, and they differ in both quantities in the scenario $k=3$. See also Figure~\ref{rec:stump} for a respective illustration. The performance of COAT is assessed in terms of its power to reject the null-hypothesis (\ref{eq:H0_both}) for the informative covariate $X_{1}$, and to uncover the correct subgroups as measured by the ARI. In this respect, the values of $\mu_{k}$ and $\sigma_{k}$ have been chosen in such a way that the power of a respective two-sample t-test would range between $0.372$ and $0.995$ for the given sample sizes \citep[]{chow2017sample}.

Finally, in the Tree Case, we again consider an outcome $Y \sim \mathcal{N}(\mu_{k}, \sigma_{k})$ with $k \in \{1,2\}$ and two informative, $X_{1}$ and $X_{2}$, and three uninformative, $X_{3}$, $X_{4}$ and $X_{5}$, standard-normally distributed covariates, resulting in three or four subgroups (see Figure~\ref{rec:tree}), according to

$$
(\mu_{k}, \sigma_{k}) = \begin{cases}
			(\mu_{1} = 0.3 \cdot I(X_{2} \geq Q_{0.75}) + 0.5 \cdot I(X_{2} < Q_{0.75}) \cdot \\ I(X_{1} \geq Q_{0.4}), \sigma_{1} = 1+I(X_{2} \geq Q_{0.75})) & \text{if $k=1$,} \\
			(\mu_{2} = 0.5 \cdot I(X_{1} \geq Q_{0.4}), \sigma_{2} = 1+I(X_{2} \geq Q_{0.6})) & \text{if $k=2$}.
		 \end{cases}
$$

The values of $\mu_{k}$ and $\sigma_{k}$ in scenario $k = 1$ have been chosen such that it offers a first split with respect to $\sigma_{1}^{2}=\mathrm{Var(Y|X)}$, which deviates between the subgroups defined by the split point $Q_{0.75}$ in $X_{2}$, while $\mu_{1}$ takes the same value $0.4 \cdot 0+0.6 \cdot 0.5=0.3$ on both sides of this split point. Subsequently, a second split could be performed with respect to $\mu_{1}=\mathbb{E}(Y|X)$ as it differs between the subgroups defined by the split point $Q_{0.4}$ in $X_{1}$ where $X_{2} < Q_{0.75}$. In the second scenario, the split point $Q_{0.6}$ in $X_{2}$ defines a split with respect to $\sigma_{2}^{2}=\mathrm{Var(Y|X)}$, and the split point $Q_{0.4}$ in $X_{1}$ defines a split with respect to $\mu_{2}=\mathbb{E}(Y|X)$, resulting in four subgroups (see Figure~\ref{rec:tree}).

\setlength{\parindent}{3ex}
\subsection{Results} \label{sim:results}
\begin{figure}
    \centering
    \includegraphics[width=0.6\textwidth]{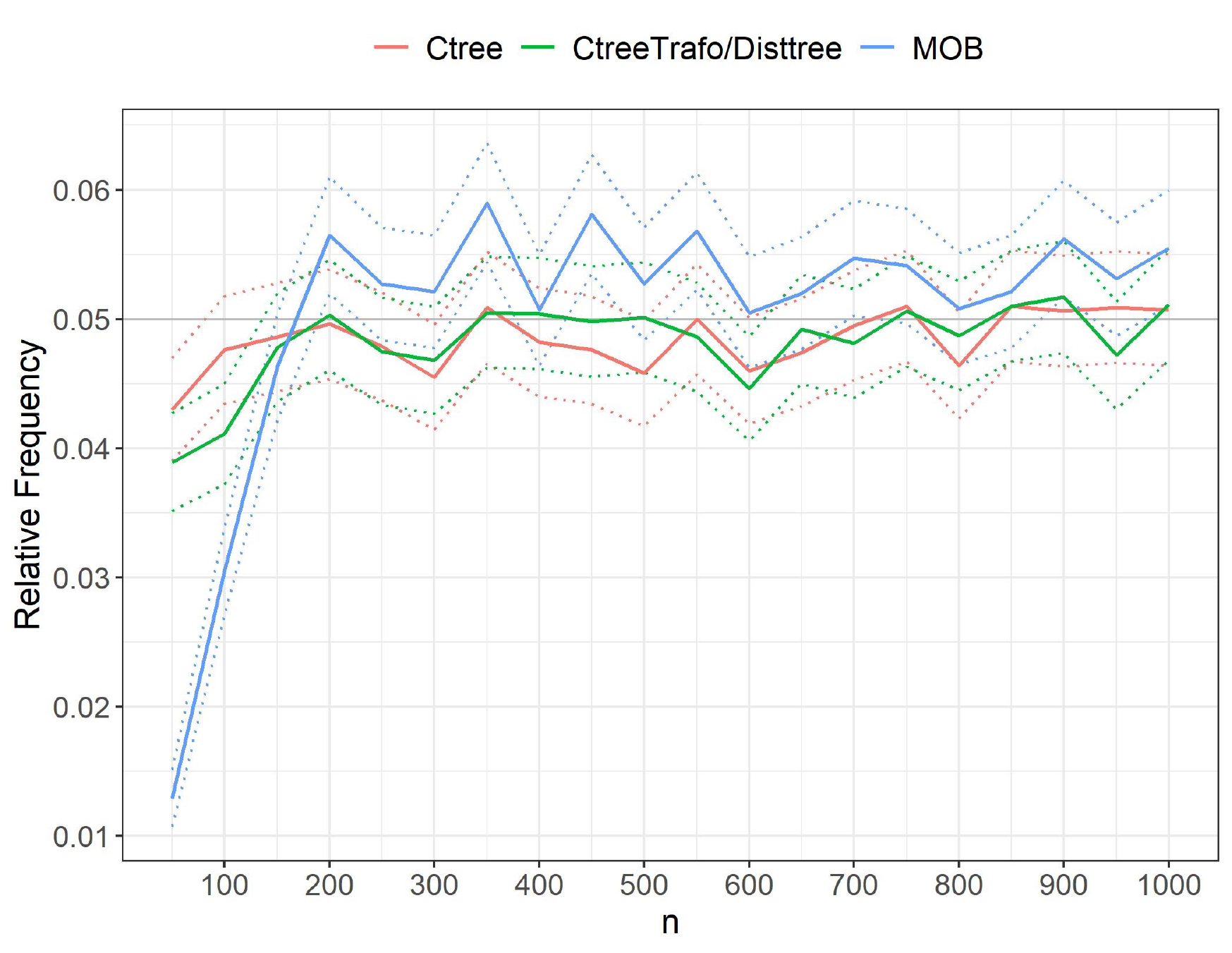}
    \caption{Relative frequency of statistically significant p-values observed in the root nodes of the COAT models fit to data of increasing sample size in the Null Case with 10000 replications. These estimates of the type-I error probability are presented with pointwise $95\%$ confidence intervals (dashed lines).}
    \label{Eval1}
\end{figure}

We first investigate the estimated type-I error probabilities of COAT in dependence of sample size in the Null Case. CTree and COAT by CTreeTrafo/DistTree show similar performance with relative rejection frequencies of the null-hypothesis reaching from $3.8\%$ to $5.5\%$, which are close to the nominal significance level of $0.05$ and appear to be independent of sample size (Figure~\ref{Eval1}). Please note that CTree only tests the first null-hypothesis in (\ref{eq:H0_each}) while COAT by CTreeTrafo/DistTree tests the null-hypothesis (\ref{eq:H0_both}). On the contrary, the COAT implementation by MOB does not seem to exploit the nominal significance level of $0.05$ well for smaller sample sizes as it rejects the null-hypothesis (\ref{eq:H0_both}) in only $1.3\%$ and $3.3\%$ of the simulated cases with $n \leq 100$. With larger sample sizes of $n \geq 200$, it showed relative frequencies for the type-I error between $5.1\%$ and $5.8\%$, which are slightly but clearly increased beyond the nominal significance level of $0.05$.

\begin{figure}
    \centering
    \begin{subfigure}[b]{0.5\textwidth}
    \includegraphics[width=0.8\textwidth,valign=c]{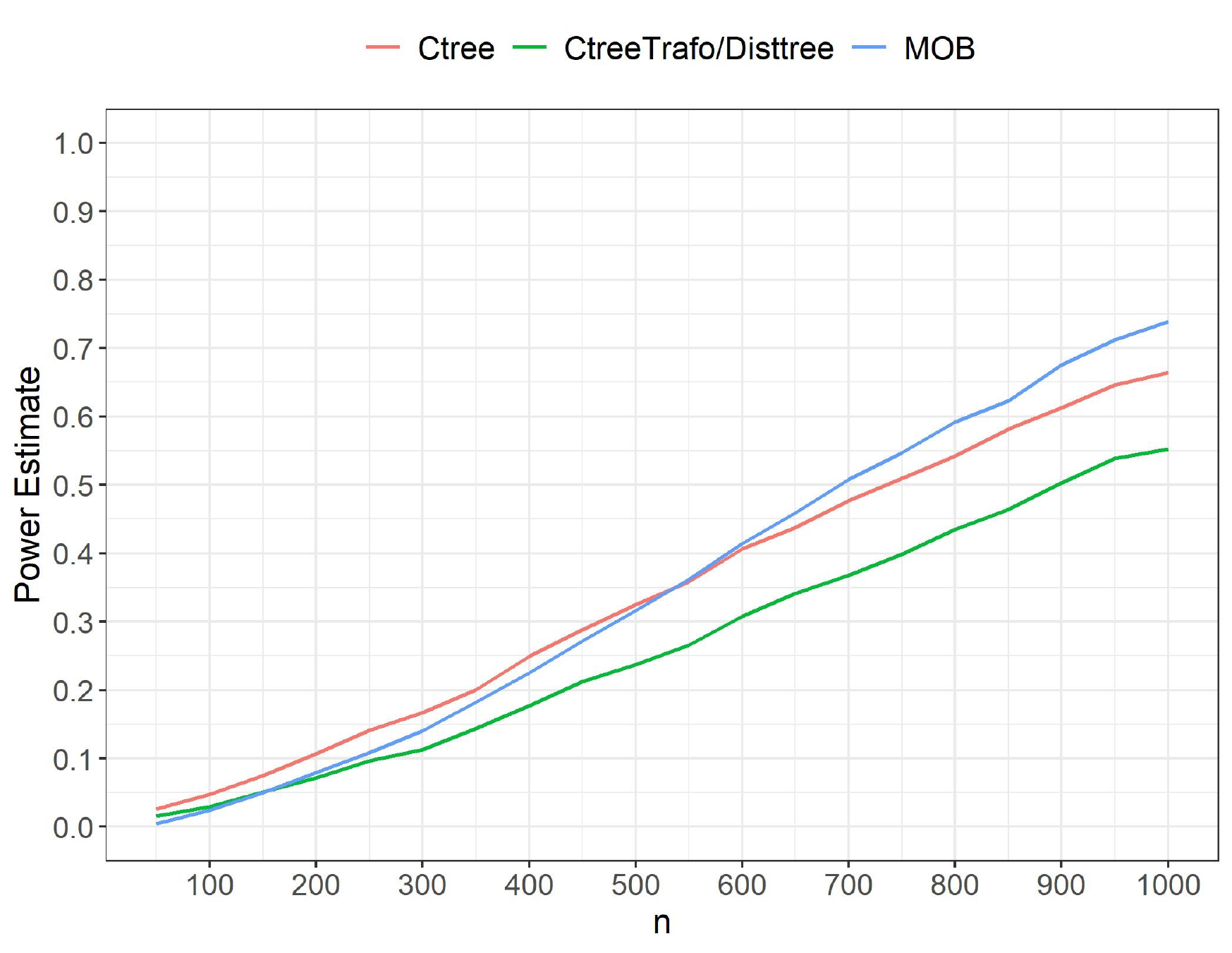}
    \subcaption[]{$k=1$}
    \label{fig:sel1}
    \end{subfigure}
    \begin{subfigure}[b]{0.35\textwidth}
    \includegraphics[width=1.15\textwidth,valign=c]{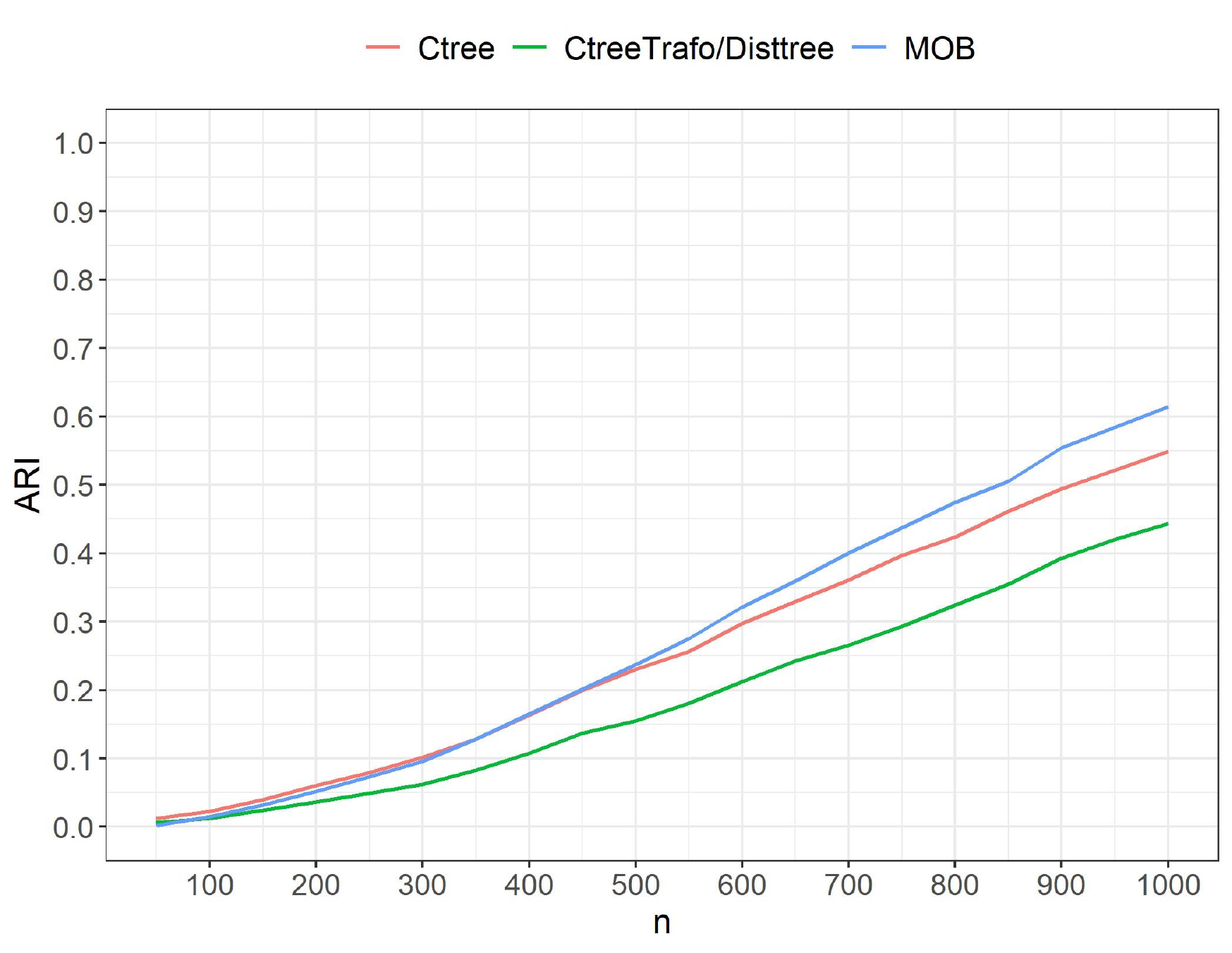}
    \subcaption[]{$k=1$}
    \label{fig:ARI1}
    \end{subfigure}
    \begin{subfigure}[b]{0.5\textwidth}
    \includegraphics[width=0.8\textwidth,valign=c]{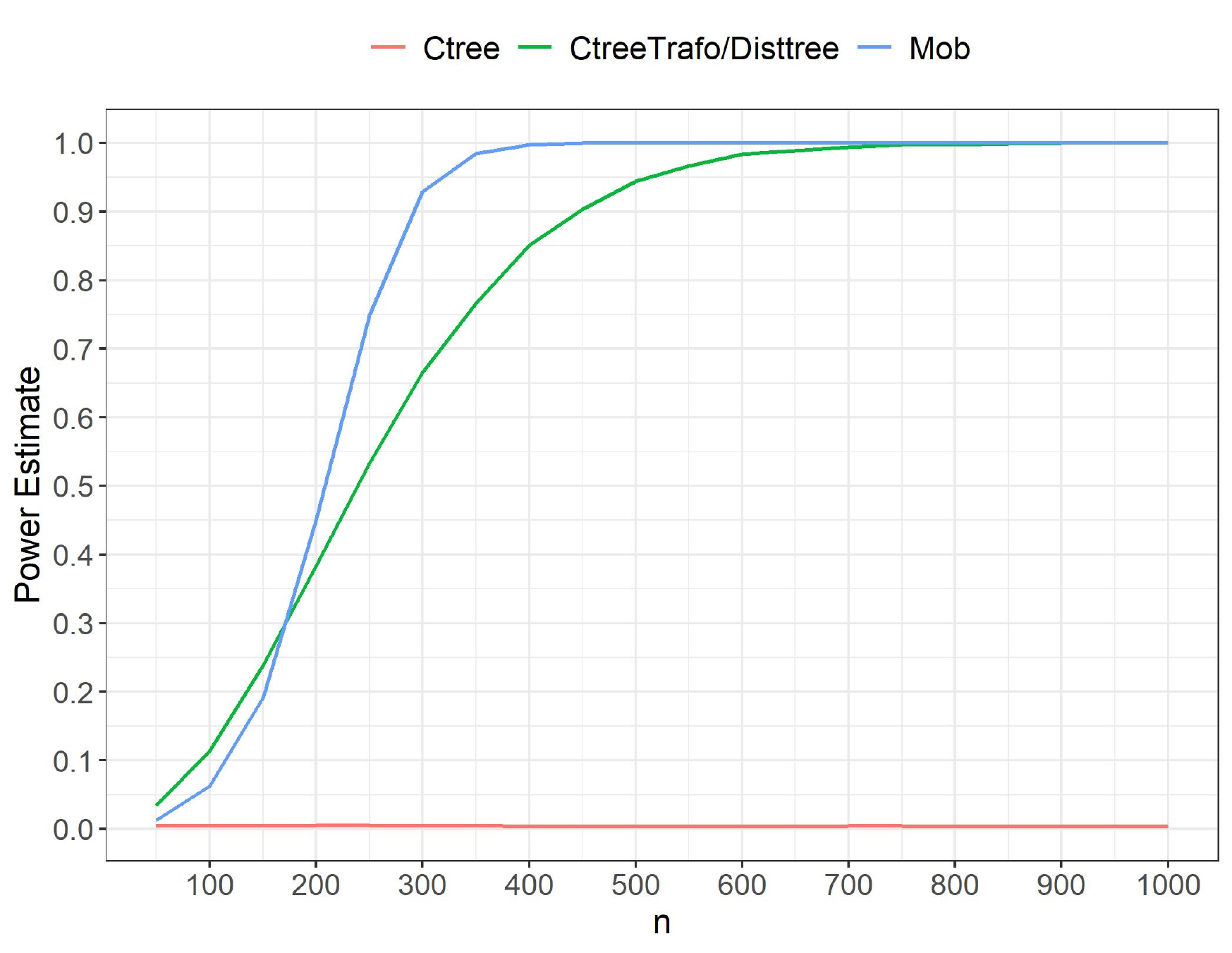}
    \subcaption[]{$k=2$}
    \label{fig:sel2}
    \end{subfigure}
    \begin{subfigure}[b]{0.35\textwidth}
    \includegraphics[width=1.15\textwidth,valign=c]{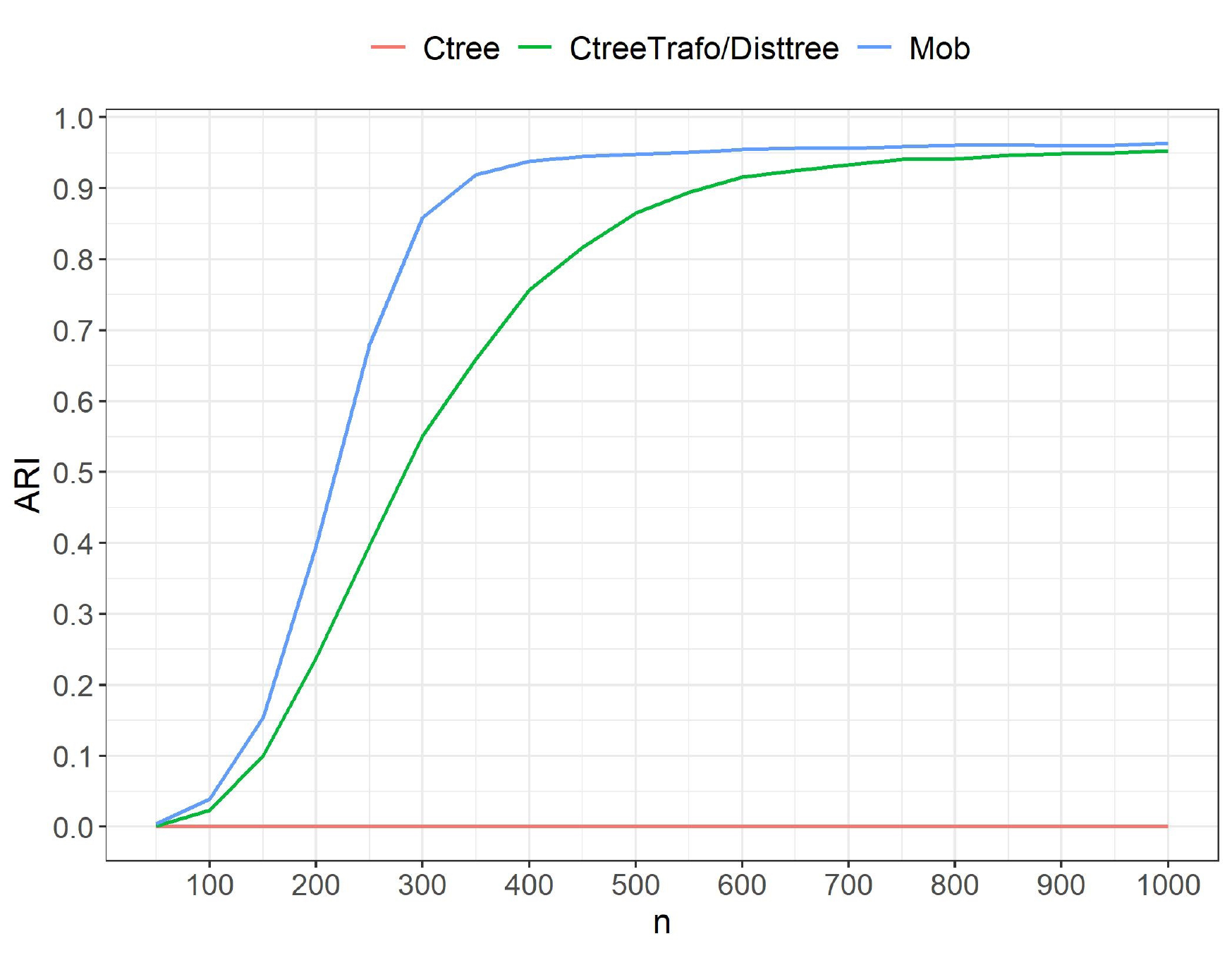}
    \subcaption[]{$k=2$}
    \label{fig:ARI2}
    \end{subfigure}
    \begin{subfigure}[b]{0.5\textwidth}
    \includegraphics[width=0.8\textwidth,valign=c]{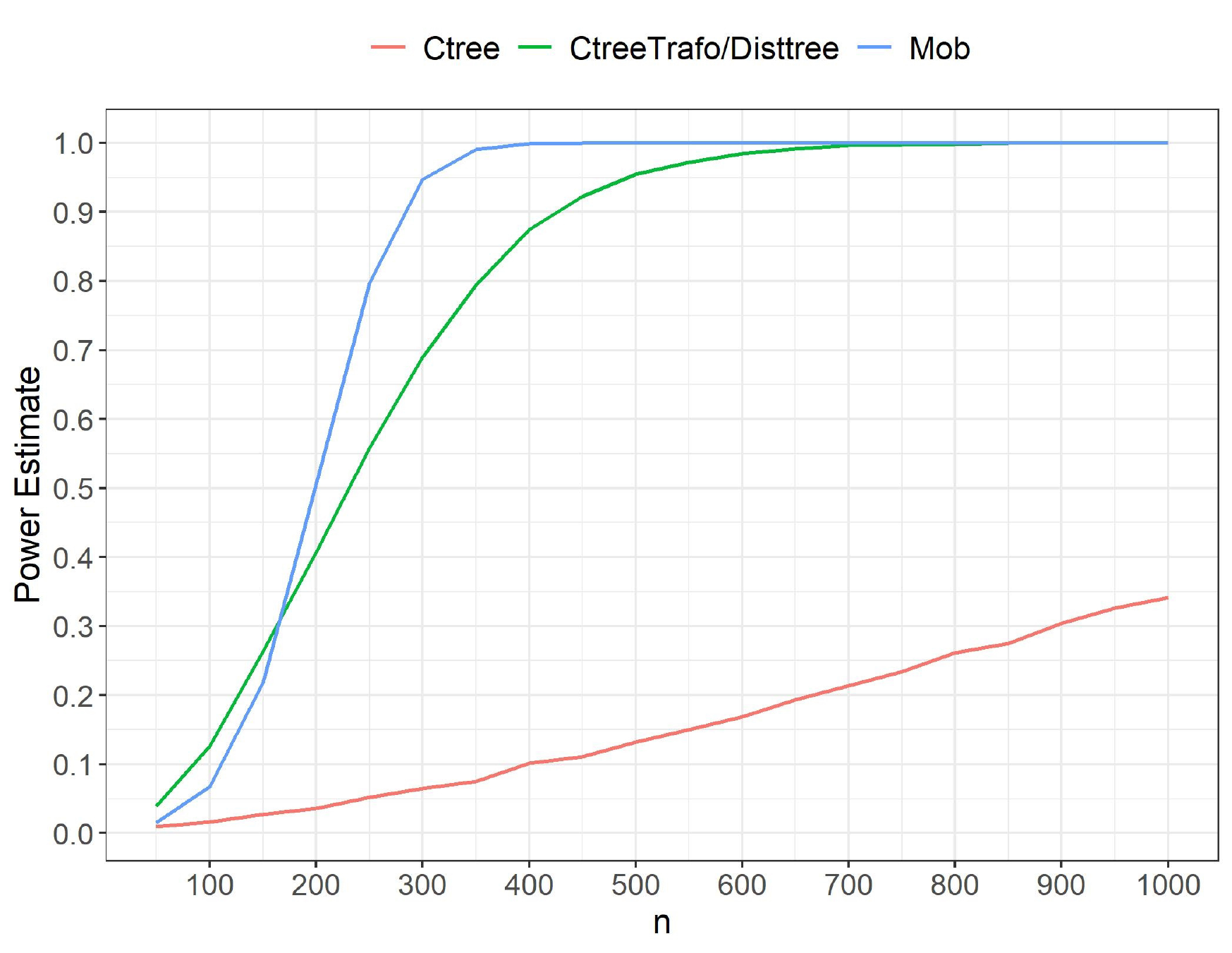}
    \subcaption[]{$k=3$}
    \label{fig:sel3}
    \end{subfigure}
    \begin{subfigure}[b]{0.35\textwidth}
    \includegraphics[width=1.15\textwidth,valign=c]{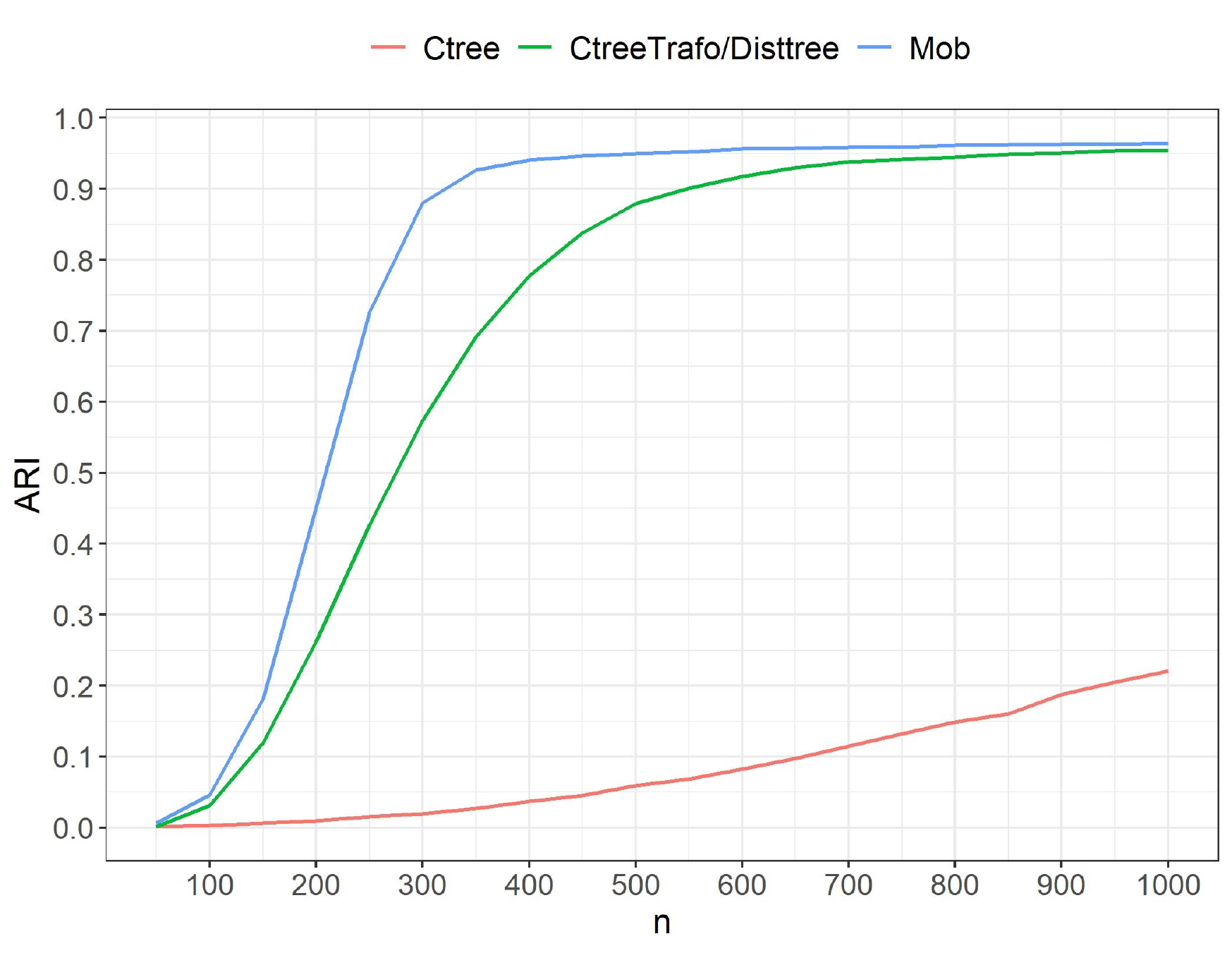}
    \subcaption[]{$k=3$}
    \label{fig:ARI3}
    \end{subfigure}
    \caption{Power estimates (\ref{fig:sel1}, \ref{fig:sel2}, \ref{fig:sel3}) and Adjusted Rand Index (ARI) (\ref{fig:ARI1}, \ref{fig:ARI2}, \ref{fig:ARI3}) for CTree, CTreeTrafo, DistTree and MOB in the three Stump Case scenarios $k \in \{1,2,3\}$, for increasing sample size. The maximum width of the pointwise $95\%$ confidence intervals was only $1.97\%$, which is why they are not presented in the plots.}
    \label{Eval2}
\end{figure}

The performance of COAT in the Stump Case in terms of the power to reject the null-hypothesis (\ref{eq:H0_both}) for the informative covariate $X_{1}$ is estimated by the respective relative frequencies of the association of $X_{1}$ and the outcome being significant at the $5\%$ level in the root node of the tree models (Figure~\ref{Eval2}). When only the expectation $\mu_{1}$ but not the variance $\sigma_{1}^{2}$ varies between the defined subgroups (i.e. scenario $k=1$), CTree and MOB perform best. However, for the case where only the variance $\sigma_{1}^{2}$ varies (i.e. scenario $k=2$), the performance of CTree decreases as it has not been enabled through a respective definition of the transformation function $h(\cdot)$ to detect such variation. Again, the MOB tree performs best, closely followed by CTreeTrafo and DistTree.

\begin{figure}
\centering
    \begin{subfigure}[b]{0.5\textwidth}
    \includegraphics[width=0.8\textwidth,valign=c]{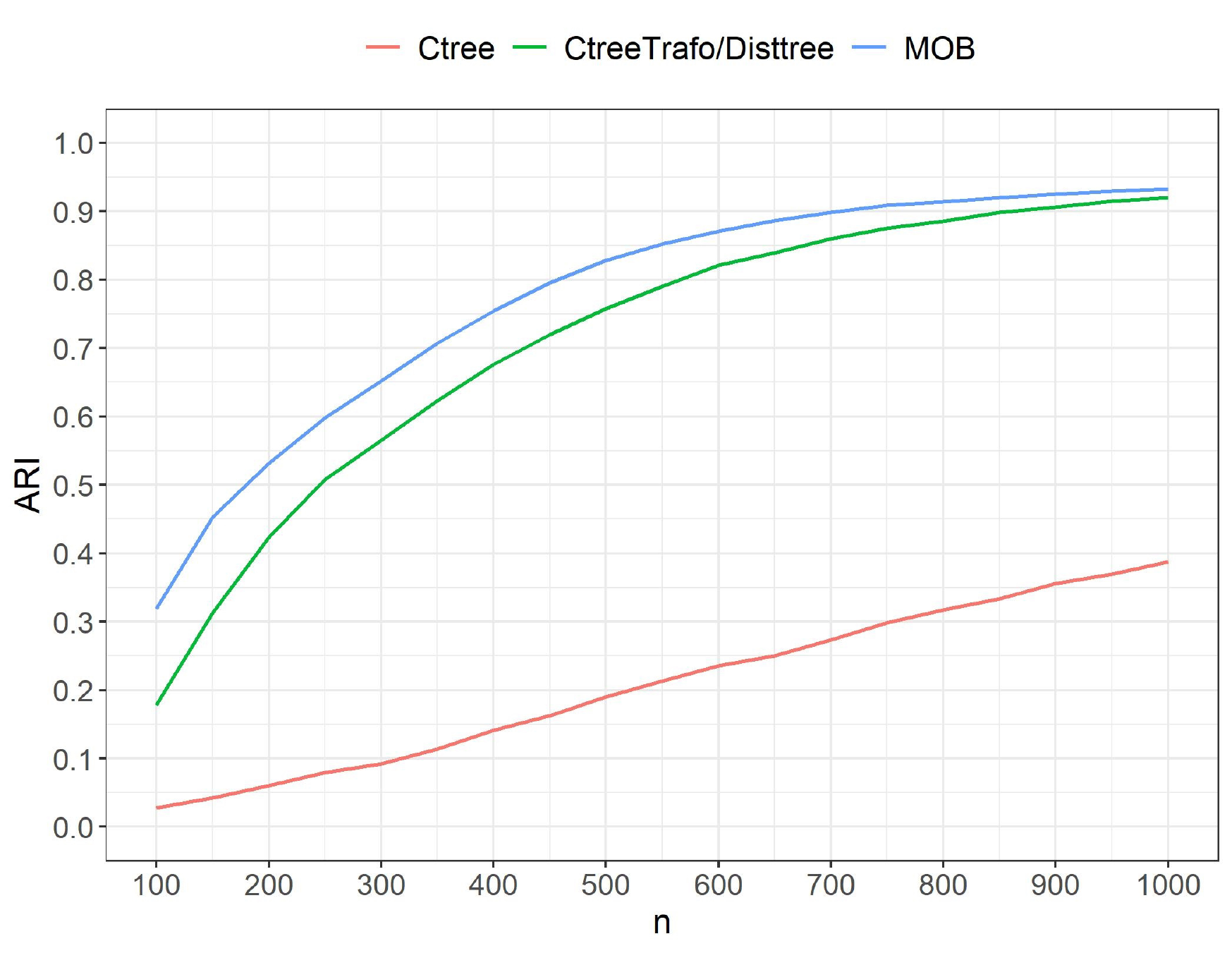}
     \subcaption[]{$k=1$}
    \end{subfigure}
    \begin{subfigure}[b]{0.35\textwidth}
    \includegraphics[width=1.15\textwidth,valign=c]{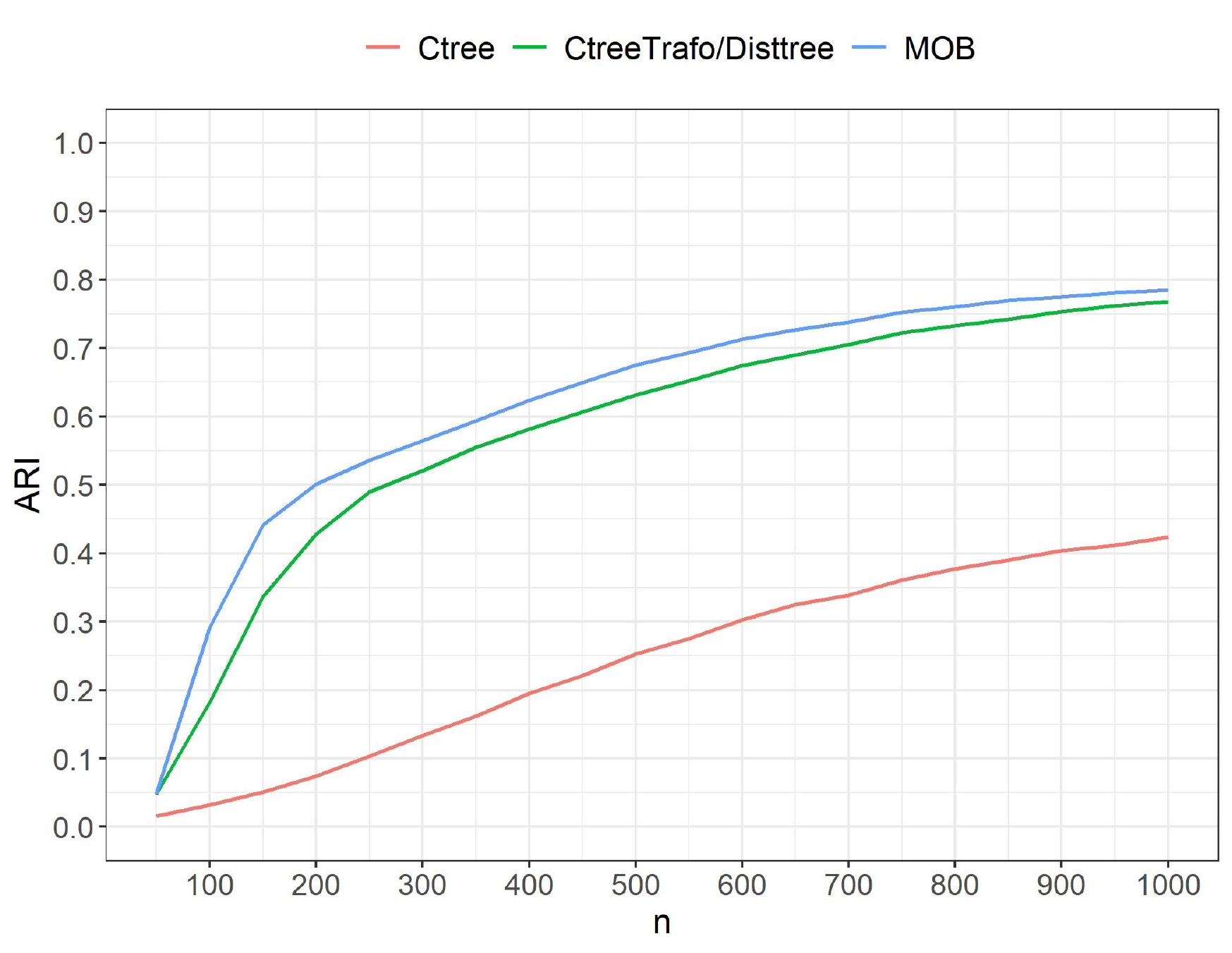}
     \subcaption[]{$k=2$}
    \end{subfigure}
    \caption{Adjusted Rand Index (ARI) of CTree, CTreeTrafo, DistTree and MOB in the two Tree Case scenarios \textbf{$k \in \{1,2\}$}, for increasing sample size. The maximum width of the pointwise $95\%$ confidence intervals was only $0.96\%$, which is why they are not presented in the plots.}
    \label{Eval4}
\end{figure}

\begin{figure}[H]
    \centering
    \includegraphics[width=\textwidth,valign=c]{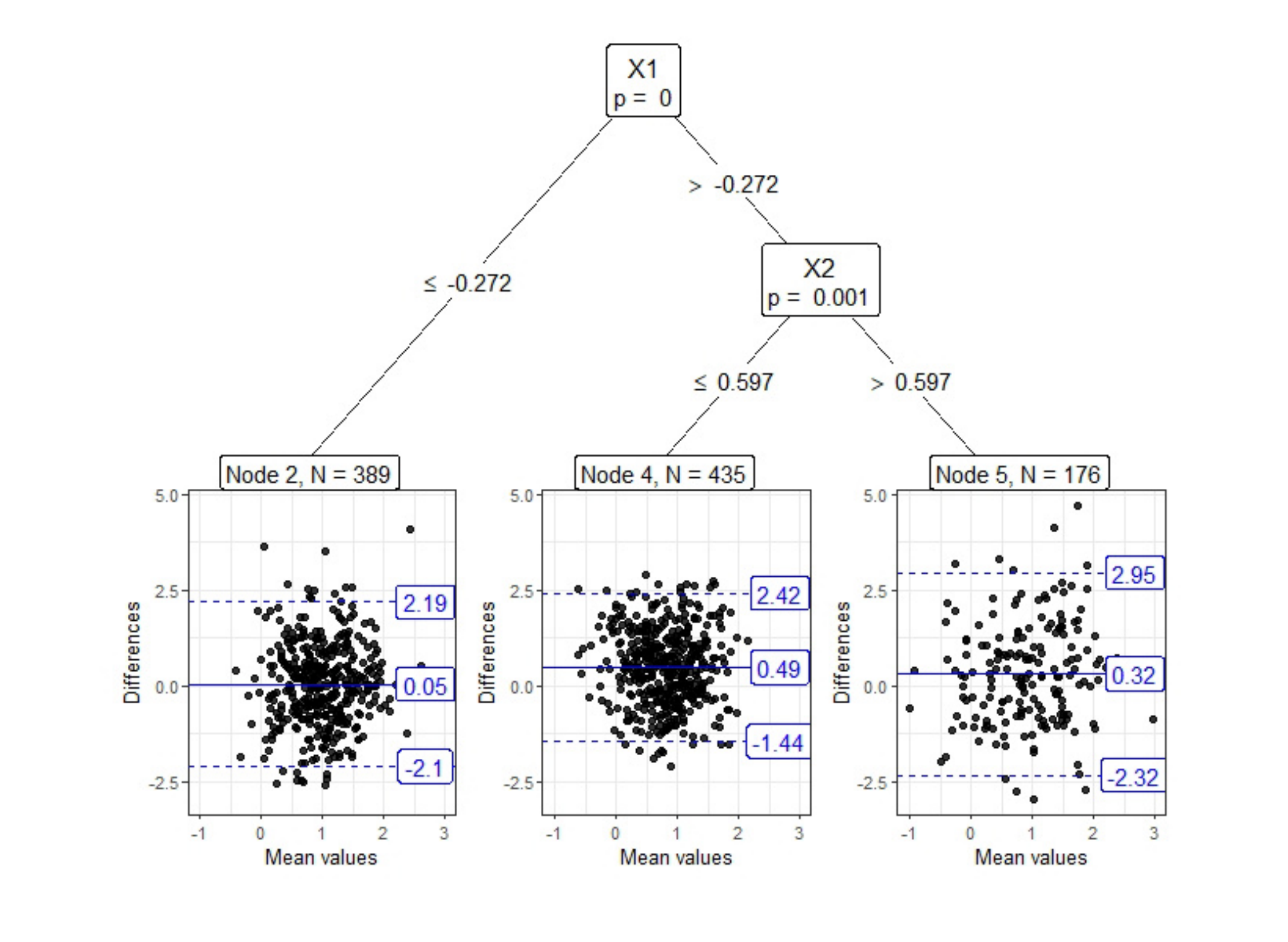}
    \caption{COAT by CTreeTrafo for conditional agreement of simulated data in the Tree Case scenario $k=1$.}
    \label{fig:example}
\end{figure}

However, power estimates do not indicate whether the true subgroups are correctly specified. Therefore, the ARI has also been investigated for the Stump Case and the Tree Case. The average ARI is plotted against increasing sample size in Figure~\ref{Eval2} and in Figure~\ref{Eval4}, respectively. As expected, the ARI increases as the sample size increases in both cases. However, Ctree can only keep up with the COAT implementations when there is only variation in the expectation $\mu_{k}$ and not in the variance $\sigma_{k}$. COAT seems to be able to cope even with the more complex setting when there are more than two true subgroups. Overall, the results for estimated power and ARI are largely comparable and lead to identical conclusions regarding the performance of the modeling approaches. An example of a tree case is given in Figure~\ref{fig:example}.

\section{Application example}
\label{sec:data}
To demonstrate the applicability and relevance of COAT in research, it is applied to a real data example of $50$ study participants who wore different accelerometers, namely one ActiGraph and two Actiheart, simultaneously for $24$ hours\citep[]{henriksen2019validity}. The ActiGraph was placed on their right hip, one Actiheart was placed in the upper position of chest, and the second Actiheart in the lower position. Both accelerometers are considered valid for estimating activity energy expenditure (AEE). The difference is that the Actiheart reports it directly, while the ActiGraph uses both uniaxial and triaxial activity counts to calculate it. In the present application example, the agreement of daily measurements of AEE (kilocalories) is compared between different pairs of two accelerometers each, conditional on the participants' age, sex, height, and weight. As described in Section~\ref{sec:coat}, we also include the mean AEE measurements along with the other covariates as a potential explanatory variable. Two cases with missing values were removed from the data. Characteristics of the participants are presented in Table \ref{tab:part_char}.

\begin{table}
    \centering
    \begin{tabular}{ll}
    \hline
    Variables & n(\%); Median (IQR) \\
    \hline
    Female & $24 \ (50\%)$ \\
    Age (years) & $40 \ (35, 57)$ \\
    Height (cm) & $174 \ (166, 182)$ \\
    Weight (kg) & $75 \ (63, 86)$ \\
    \hline
    \end{tabular}
    \caption{Participant characteristics of the application study ($n=48$).}
    \label{tab:part_char}
\end{table}

\begin{figure}
    \centering
    \begin{subfigure}[b]{0.9\textwidth}
      \includegraphics[width=\textwidth,valign=c]{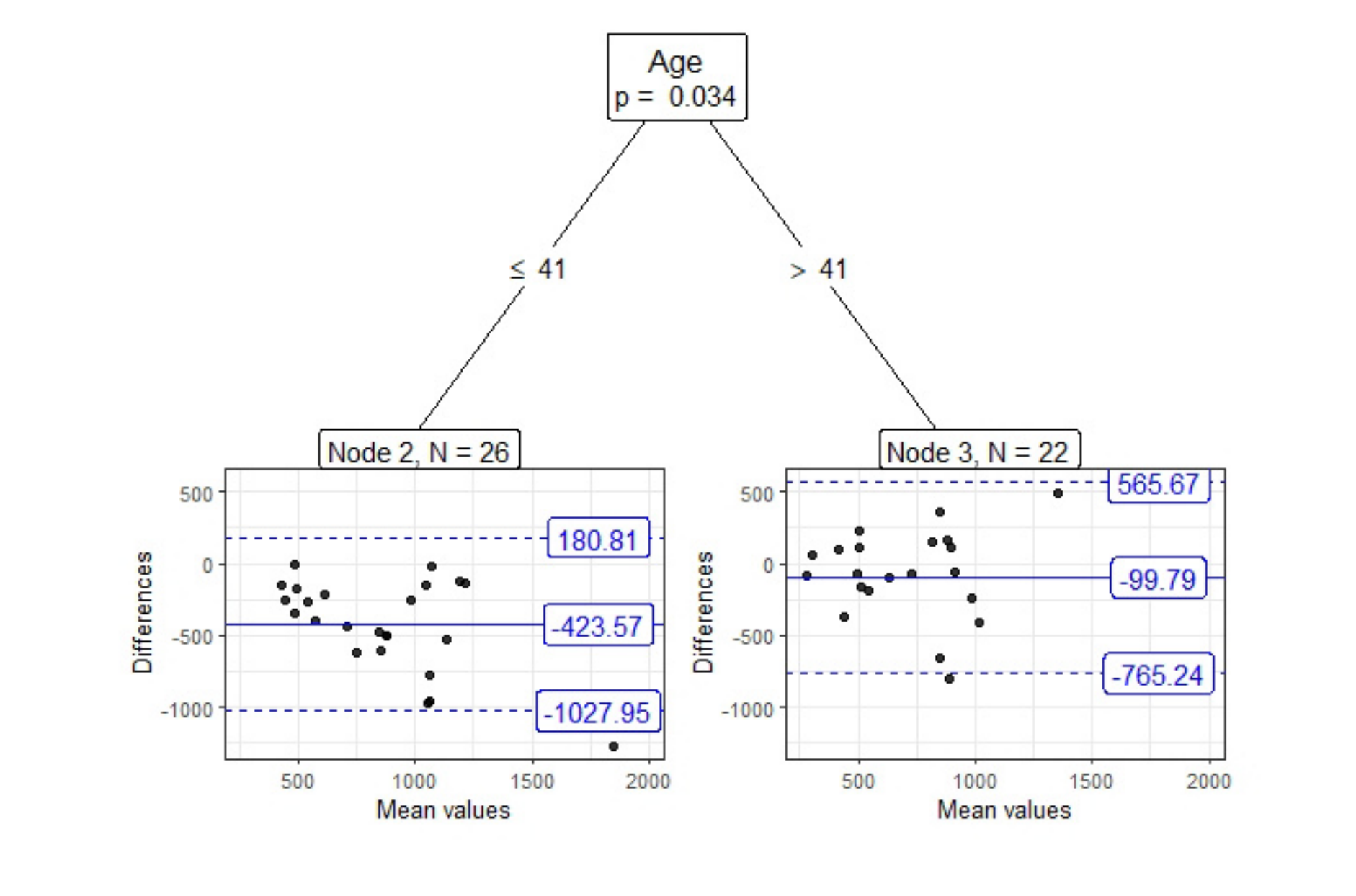}
      \subcaption{COAT by MOB.}
      \label{COAT1}
    \end{subfigure}
    \begin{subfigure}[b]{0.9\textwidth}
      \includegraphics[width=\textwidth,valign=c]{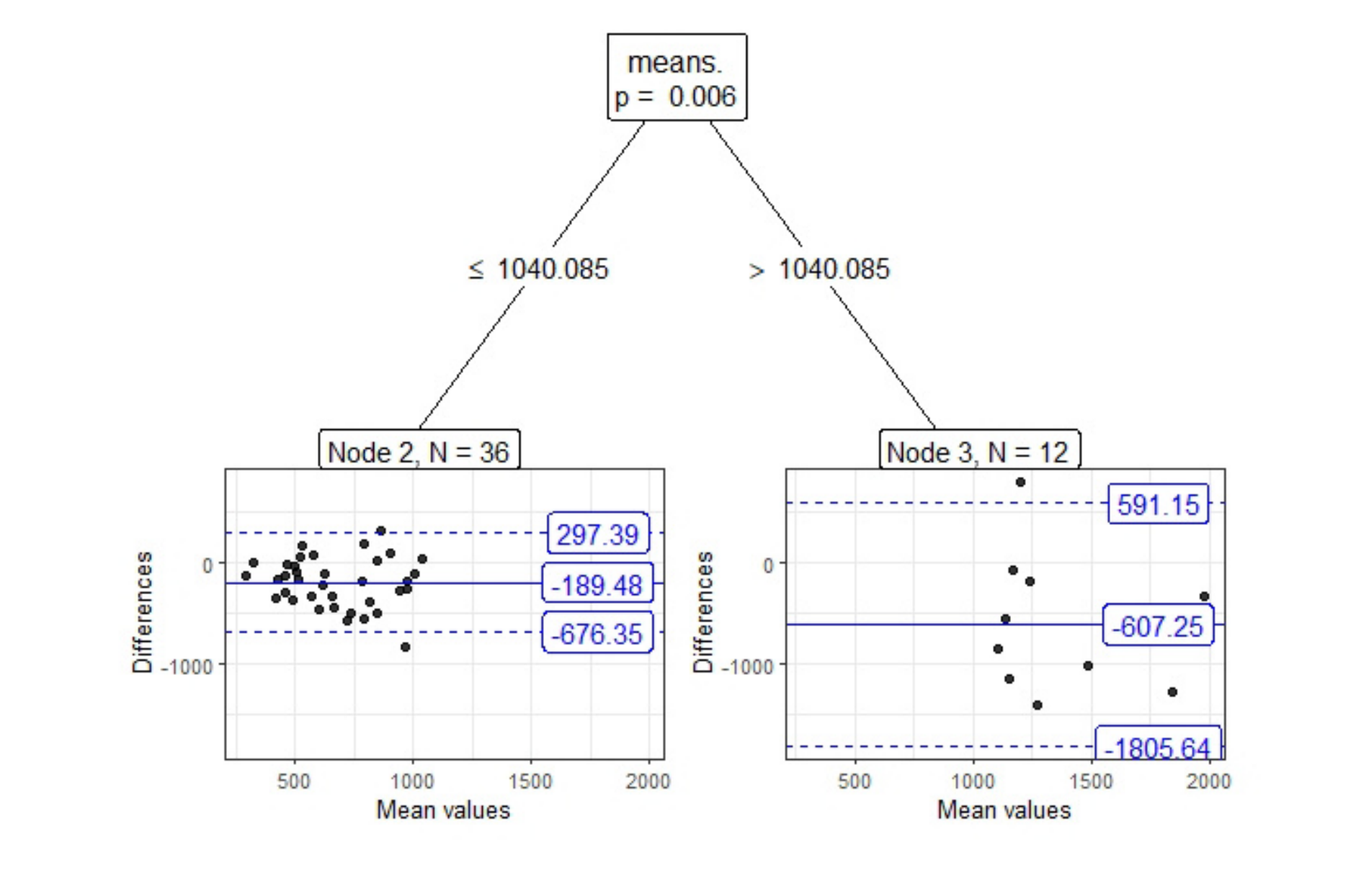}
      \subcaption{COAT by CTreeTrafo.}
      \label{COAT2}
    \end{subfigure}
    \caption{COAT for conditional agreement of activity energy expenditure (AEE) measurements of two accelerometers. Note that different pairs of accelerometers are compared in (\ref{COAT1}, ActiGraph based on triaxial activity counts and Actiheart in upper position) and (\ref{COAT2}, ActiGraph based on triaxial activity counts and Actiheart in lower position). See Section~\ref{sec:data} for details.}
    \label{COAT}
\end{figure}

Figure~\ref{COAT1} shows that for one pair of compared accelerometers, COAT by MOB is able to identify subgroups of participants, which are heterogeneous regarding the bias and width of LoA depending on age ($p=0.034$). Better agreement, in terms of bias decreasing by about $324$ kilocalories, is obtained for patients older than $41$ years. With two other accelerometers, COAT by CTreeTrafo showed that agreement may be conditional on the magnitude of measurements (Figure~\ref{COAT2}). With an average AEE $> 1040$ kilocalories, the bias in agreement increases by about $220$ kilocalories and the width of the LoA increases by about $444$ kilocalories ($p=0.006$).

\section{A Bland-Altman test} \label{BA:test}
\begin{figure}[H]
\centering
    \includegraphics[width=0.8\textwidth]{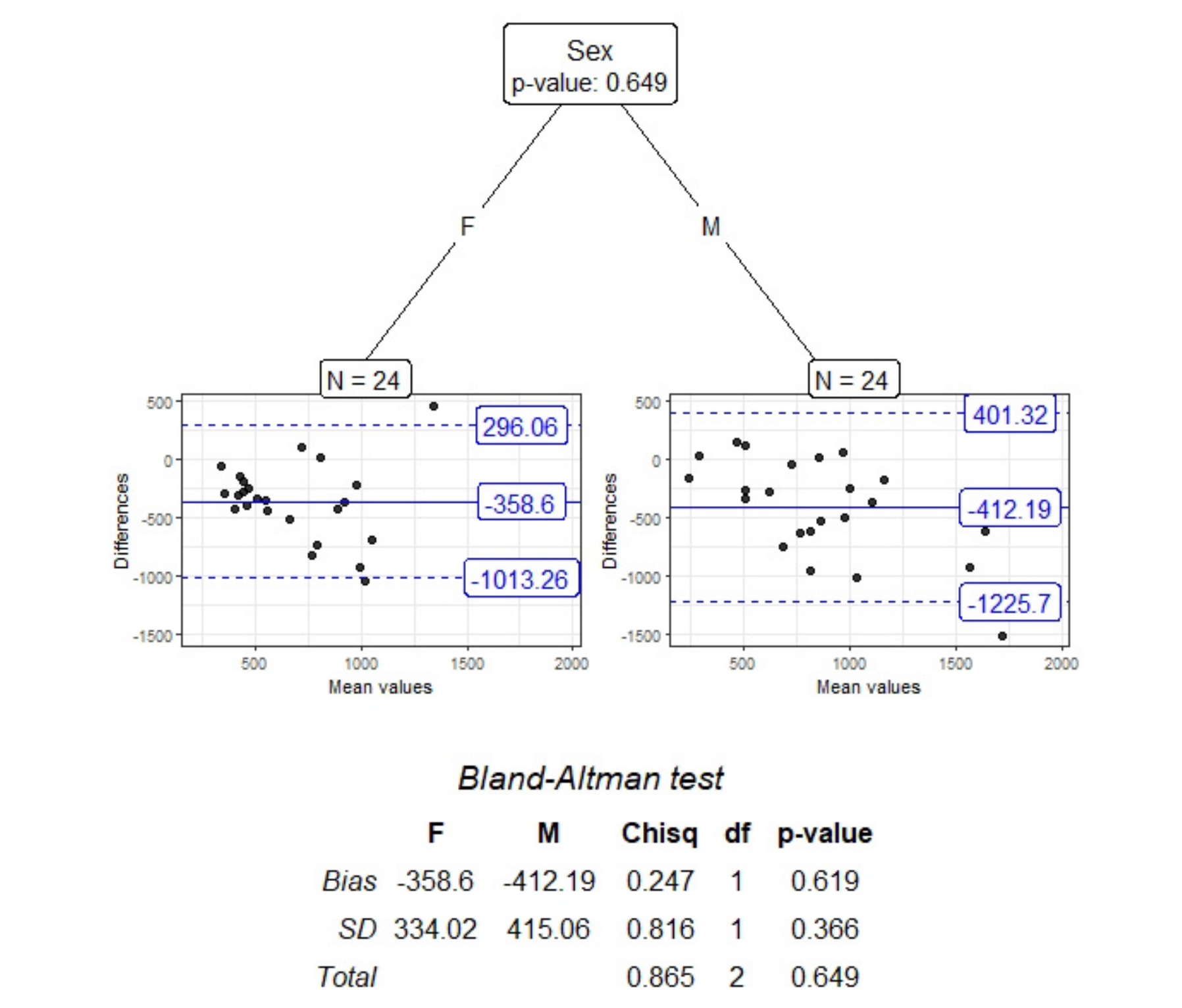}
    \caption{Bland-Altman (BA) test of difference in method agreement of activity energy expenditure (AEE) measurements between female (F) and male (M) participants in the application study. ActiGraph based on uniaxial activity counts and Actiheart in the upper position are compared.}
    \label{COAT3}
\end{figure}

It has been proposed in Section~\ref{sec:coat} to apply COAT to perform a two-sample `Bland-Altman test' of the null-hypotheses (\ref{eq:H0_both}) and (\ref{eq:H0_each}) for comparison of agreement between (pre)defined subgroups. For example, in the application example of the previous Section~\ref{sec:data}, a researcher may be interested in a potential difference of agreement between the sexes. Figure~\ref{COAT3} shows the result of COAT by CTree, when a stump tree is generated for sex as the only covariate. In this implementation of COAT, the $\chi^2$ test statistics $c_{quad}$, degrees of freedom and respective p-values (cf. Section~\ref{sec:ctree}) are presented for testing the null-hypothesis (\ref{eq:H0_both}) and each of the null-hypotheses (\ref{eq:H0_each}) concerning differences in bias and width of LoA between the considered subgroups. Corresponding estimates of $\mathbb{E}(Y|X)$ and $\mathrm{Var(Y|X)}$ are provided for each subgroup, too. In the present case, no statistically significant difference was found between the sexes in terms of bias ($p=0.619$), the width of the LoA ($p=0.366$) and both of these quantities ($p=0.649$). Please note that these three p-values are not adjusted for the multiple testing problem, but are easily suitable for conducting a sequential test procedure (starting with the test of both quantities, followed by the test of the individual quantities). Other corrections, such as the Bonferroni correction, are of course also possible. Similarly to the proposed approach, a Bland-Altman test could also be used for a single predictor variable of any scale, for example if there is no definition of two subgroups. However, this case is already covered by COAT, as described above.

\section{Discussion}
The contribution of the present work to the field of method comparison studies is fourfold. First, the concept of conditional method agreement is introduced and formalized. Second, respective statistical modeling by recursive partitioning is proposed introducing conditional method agreement trees (COAT). Third, a respective Bland-Altman test is suggested to test for differences in agreement, with respect to the bias and width of LoA, between (pre)defined subgroups. Fourth, COAT is made publicly available through the R package \texttt{coat}.

COAT provides a solution to simultaneously address the research questions of method agreement and potential dependence on covariates in a unifying framework. It therefore exploits the fact that conditional method agreement can be parameterized through the expectation $\mathbb{E}(Y|X)$ and variance $\mathrm{Var(Y|X)}$ of paired differences between two methods' measurements. Correctly specified tree-based models are used for estimation of these conditional parameters and enable the definition of subgroups with different agreement.

Results of the simulation study indicate that the implementations of COAT by CTree (i.e. CTreeTrafo) and DistTree are able to control the type-I error probability at the nominal significance level, independent of sample size. By contrast, the implementation by MOB showed a decisively decreased error rate with small sample sizes and a slightly increased error rate with larger sample sizes. Therefore, it cannot be recommended for COAT in its present form, and further research could be directed towards robust variance estimation and improvements in distributional approximations for possible correction. All implementations of COAT performed well in detecting existent subgroups with increasing sample size. The comparison to the default specification of the CTree algorithm shows that CTree without the proposed transformation only captures differences in the bias, that is in the conditional expectation $\mathbb{E}(Y|X)$, but cannot uncover differences in the width of LoA, that is in the variance $\mathrm{Var(Y|X)}$.

Observed differences between the implementations of COAT arise from the testing strategy. Both CTreeTrafo and DistTree compute quadratic test statistics which are equivalent, as has been analytically shown in appendix \ref{TestStat}. In this respect, DistTree can be considered a special case of CTree with the appropriate transformation function $h(\cdot)$ as defined in Section~\ref{sec:ctree}. By contrast, MOB is based on fluctuation tests for parameter instability in regression model fits. 

The application study exemplifies the potential of COAT for medical research. Therefore, subgroups with heterogeneous method agreement in activity energy expenditure (AEE) measurements could be identified in terms of bias and width of LoA depending on covariates and the size of the AEE measurements. From the perspective of the applicant, which could be a manufacturer of accelerometers, a researcher, investigator or treating physician, one can then recommend which accelerometer to use or how to improve measurements in a particular setting for a particular person.

It should be noted that the results of COAT are exploratory, unless it is used to conduct a two-sample Bland-Altman test of different agreement between (pre)defined subgroups. In the latter case it can be used for confirmatory hypothesis testing. In this context, it should also be mentioned that CTree, DistTree and MOB by default apply a Bonferroni correction to the multiple testing problem that occurs when a test-based splitting is performed based on multiple covariates. At present, COAT is limited to the case of single measurements per observational unit or subject. A modification for repeated measurements is currently being developed.

\section{Conclusion}
COAT enables the analysis of method agreement in dependence of covariates and mean measurements by conditional modeling and exploratory or confirmatory hypothesis testing. It is made publicly available through the R package \texttt{coat}.

\section*{Data availability}
Data underlying the application study may be obtained from the authors of the original study upon reasonable request \citep[]{henriksen2019validity}. The code of the simulation studies is provided as supplementary material. COAT is made publicly available through the associated R package \texttt{coat} on the Comprehensive R Archive Network (CRAN).

\section*{Conflicts of interest}
None to declare.

\section*{Funding}
This study was funded by the Deutsche Forschnugsgemeinschaft (DFG, German Research Foundation) - Projektnummer (grant number) $447467169$.

\section*{Author contributions}
S.K. and A.H. drafted the manuscript, performed the statistical analyses and interpreted the results. André H. extracted and prepared the data used in the application study. All authors revised the manuscript for its content and approved the submission of the final manuscript.

\section*{Acknowledgements}
We thank Alexander Horsch for fruitful discussions and recommendations.

\section*{R Code}
R code of the performed simulation and application studies is provided as supplementary material.
The associated R package \texttt{coat} is available from GitHub.

\bibliographystyle{unsrt}
\bibliography{myreferences}

\newpage
\appendix
\section{Appendix} \label{App}

\subsection{Equality of test statistics of CTreeTrafo and DistTree} \label{TestStat}
In this section, we analytically show that the test statistics $c_{quad}(\cdot)$ of CTreeTrafo and DistTree are equivalent for the case of numeric split variables. For a clearer and more comprehensible presentation of the already very extensive proof, the slightly more complex case of categorical variables has been omitted. However, it can be shown analogously. Recall the test statistic used for CTreeTrafo and DistTree:
\begin{equation} \label{eq:teststat}
    c_{quad}(t_j, \mu_j, \Sigma_j) = (t_j-\mu_j) \Sigma_j^+ (t_j-\mu_j)^\top.
\end{equation}

In the following, we define each element $t_j$, $\mu_j$ and $\Sigma_j$ in (\ref{eq:teststat}) based on the formulas from the original publication \citep[]{strasser1999asymptotic} and as outlined in sections \ref{sec:ctree} and \ref{sec:distttree}. To simplify notation, we omit the index $j$, which specifies a particular split variable. The weights $\omega_i$ are chosen to be $1$ focusing on the observations of a given node in a tree. $\Sigma^+$ is in our case equivalent to $\Sigma^{-1}$.

\subsubsection*{CTreeTrafo}

In CTreeTrafo the statistic

\[
 \begin{split}
    t & = vec \left( \sum\limits_{i=1}^n \underbrace{\omega_i}_{=1} \underbrace{g(x_i)}_{=x_{i}} (y_i, (y_i - \overline{y})^2)^\top  \right) = \sum\limits_{i=1}^n x_i (y_i, (y_i - \overline{y})^2)^\top \\
      & = \left(\sum\limits_{i=1}^n x_i y_i, \sum\limits_{i=1}^n x_i \underbrace{(y_i - \overline{y})^2}_{=s_i}\right)^\top =  \left(\sum\limits_{i=1}^n x_i y_i, \sum\limits_{i=1}^n x_i s_i \right)^\top
 \end{split}
\]

\noindent has the expectation

\[
 \begin{split}
    \mu & = vec \left( \left(\sum\limits_{i=1}^n \omega_i g(x_i)     \right) \frac{1}{n} \sum\limits_{i=1}^n \omega_i (y_i, (y_i - \overline{y})^2)^\top  \right) = \left( \sum\limits_{i=1}^n x_i \right) \frac{1}{n} \sum\limits_{i=1}^n (y_i, (y_i - \overline{y})^2)^\top   \\
      & = \left( \frac{1}{n} \sum\limits_{i=1}^n x_i \sum\limits_{i=1}^n y_i, \frac{1}{n} \sum\limits_{i=1}^n x_i \sum\limits_{i=1}^n (y_i - \overline{y})^2 \right)^{T} = \left( n \overline{x} \overline{y}, \overline{x} \sum\limits_{i=1}^n (y_i - \overline{y})^2 \right)^{T} \\
      & = \left( n \overline{x} \overline{y}, n \overline{x} \overline{s} \right)^{T}
 \end{split}
\]

\noindent and covariance

\begin{equation} \label{eq3}
 \begin{split}
    \Sigma & = \frac{n}{n-1} V \otimes \left( \sum\limits_{i=1}^n \omega_i g(x_i) \otimes \omega_i g(x_i)^\top \right) - \frac{1}{n-1} V \otimes \left( \sum\limits_{i=1}^n \omega_i g(x_i) \right) \left( \sum\limits_{i=1}^n \omega_i g(x_i) \right)^\top \\
      & = \frac{n}{n-1} V \left( \sum\limits_{i=1}^n x_i  x_i \right) - \frac{1}{n-1} V \left( \sum\limits_{i=1}^n x_i \right) \left( \sum\limits_{i=1}^n x_i \right) \\
      & = \frac{n}{n-1} V \left( \sum\limits_{i=1}^n x_i^2 \right) - \frac{1}{n-1} V \left( \sum\limits_{i=1}^n x_i \right)^2  = \frac{n}{n-1} V \left( \sum\limits_{i=1}^n x_i^2 - \frac{1}{n} \left(\sum\limits_{i=1}^n x_i\right)^2 \right) \\
      & = \frac{n}{n-1} V \left( \sum\limits_{i=1}^n x_i^2 - n\overline{x}^2 \right),
 \end{split}
\end{equation}

\noindent where $\otimes$ is the Kronecker product and $V$ is defined as follows:

\[ 
\begin{split}
    V & = \frac{1}{n} \sum\limits_{i=1}^n \omega_i \left( (y_i, (y_i - \overline{y})^2) - \frac{1}{n} \sum\limits_{i=1}^n \omega_i (y_i, (y_i - \overline{y})^2) \right) \left( (y_i, (y_i - \overline{y})^2) - \frac{1}{n} \sum\limits_{i=1}^n \omega_i (y_i, (y_i - \overline{y})^2) \right)^\top \\
      & = \frac{1}{n} \sum\limits_{i=1}^n \left( y_i - \frac{1}{n} \sum\limits_{i=1}^n y_i, (y_i - \overline{y})^2 - \frac{1}{n} \sum\limits_{i=1}^n (y_i - \overline{y})^2 \right) \left(y_i - \frac{1}{n} \sum\limits_{i=1}^n y_i, (y_i - \overline{y})^2 - \frac{1}{n} \sum\limits_{i=1}^n (y_i - \overline{y})^2 \right)^\top \\
      & = \frac{1}{n} \sum\limits_{i=1}^n \left(y_i - \overline{y}, (y_i - \overline{y})^2 - \frac{1}{n} \sum\limits_{i=1}^n (y_i - \overline{y})^2 \right) \left(y_i - \overline{y}, (y_i - \overline{y})^2 - \frac{1}{n} \sum\limits_{i=1}^n (y_i - \overline{y})^2 \right)^\top \\
      & = \frac{1}{n} 
      \left(
      \begin{matrix}
        \sum\limits_{i=1}^n (y_i - \overline{y})^2 & \sum\limits_{i=1}^n (y_i - \overline{y})((y_i - \overline{y})^2 - \frac{1}{n} \sum\limits_{i=1}^n (y_i - \overline{y})^2) \\
        \sum\limits_{i=1}^n (y_i - \overline{y})((y_i - \overline{y})^2 - \frac{1}{n} \sum\limits_{i=1}^n (y_i - \overline{y})^2) &  \sum\limits_{i=1}^n ((y_i - \overline{y})^2 - \frac{1}{n} \sum\limits_{i=1}^n (y_i - \overline{y})^2)^2
      \end{matrix} 
      \right) \\
      & = \frac{1}{n} 
      \left(
      \begin{matrix}
       \sum\limits_{i=1}^n s_i & \sum\limits_{i=1}^n \sqrt{s_i} (s_i - \overline{s}) \\
        \sum\limits_{i=1}^n \sqrt{s_i} (s_i - \overline{s}) & \sum\limits_{i=1}^n (s_i - \overline{s})^2
      \end{matrix}
      \right)
 \end{split}
\]

\newpage
\noindent Taken together we now obtain

\[
\begin{split}
    c_{quad}(t, \mu, \Sigma) & = \left( t - \mu \right) \Sigma^{-1} \left( t - \mu \right)^\top \\
    & = \left( \sum\limits_{i=1}^n x_i y_i - n \overline{x} \overline{y}, \sum\limits_{i=1}^n x_i s_i - n \overline{x} \overline{s} \right) \Sigma^{-1} \left( \sum\limits_{i=1}^n x_i y_i - n \overline{x} \overline{y}, \sum\limits_{i=1}^n x_i s_i - n \overline{x} \overline{s} \right)^\top \\
    & = \left( \sum\limits_{i=1}^n x_i y_i - n \overline{x} \overline{y}, \sum\limits_{i=1}^n x_i s_i - n \overline{x} \overline{s} \right) \left( \frac{n}{n-1} V \left( \sum\limits_{i=1}^n x_i^2 - n\overline{x}^2 \right) \right)^{-1} \\
    & \left( \sum\limits_{i=1}^n x_i y_i - n \overline{x} \overline{y}, \sum\limits_{i=1}^n x_i s_i - n \overline{x} \overline{s} \right)^\top \\
    & = \left( \sum\limits_{i=1}^n x_i y_i - n \overline{x} \overline{y}, \sum\limits_{i=1}^n x_i s_i - n \overline{x} \overline{s} \right) \underbrace{\left( \frac{n}{n-1} \left( \sum\limits_{i=1}^n x_i^2 - n\overline{x}^2 \right) \right)^{-1}}_{:=a} V^{-1} \\
    & \left( \sum\limits_{i=1}^n x_i y_i - n \overline{x} \overline{y}, \sum\limits_{i=1}^n x_i s_i - n \overline{x} \overline{s} \right)^\top \\
    & = \left( \sum\limits_{i=1}^n x_i y_i - n \overline{x} \overline{y}, \sum\limits_{i=1}^n x_i s_i - n \overline{x} \overline{s} \right) a \underbrace{\frac{1}{\sum\limits_{i=1}^n s_i \sum\limits_{i=1}^n x_i^2 (s_i - \overline{s})^2 - (\sum\limits_{i=1}^n x_i^2 \sqrt{s_i}(s_i - \overline{s}))^2}}_{:=b} \\
    & n 
    \left(
    \begin{matrix}
       \sum\limits_{i=1}^n (s_i - \overline{s})^2 & -\sum\limits_{i=1}^n \sqrt{s_i} (s_i - \overline{s}) \\
        -\sum\limits_{i=1}^n \sqrt{s_i} (s_i - \overline{s}) & \sum\limits_{i=1}^n s_i
    \end{matrix}
    \right)
     \left( \sum\limits_{i=1}^n x_i y_i - n \overline{x} \overline{y}, \sum\limits_{i=1}^n x_i s_i - n \overline{x} \overline{s} \right)^\top \\
     & = a \cdot b \cdot n
     \left( \sum\limits_{i=1}^n x_i y_i - n \overline{x} \overline{y}, \sum\limits_{i=1}^n x_i s_i - n \overline{x} \overline{s} \right)
     \left(
     \begin{matrix}
       \sum\limits_{i=1}^n (s_i - \overline{s})^2 & -\sum\limits_{i=1}^n \sqrt{s_i} (s_i - \overline{s}) \\
        -\sum\limits_{i=1}^n \sqrt{s_i} (s_i - \overline{s}) & \sum\limits_{i=1}^n s_i
    \end{matrix}
    \right) \\
    & \left( \sum\limits_{i=1}^n x_i y_i - n \overline{x} \overline{y}, \sum\limits_{i=1}^n x_i s_i - n \overline{x} \overline{s} \right)^\top \\
    & = a \cdot b \cdot n
    \left(
    \begin{matrix}
    (\sum\limits_{i=1}^n x_i y_i - n \overline{x} \overline{y}) \sum\limits_{i=1}^n (s_i - \overline{s})^2 - (\sum\limits_{i=1}^n x_i s_i - n \overline{x} \overline{s}) \sum\limits_{i=1}^n \sqrt{s_i} (s_i - \overline{s}), \\
    - (\sum\limits_{i=1}^n x_i y_i - n \overline{x} \overline{y}) \sum\limits_{i=1}^n \sqrt{s_i} (s_i - \overline{s}) + (\sum\limits_{i=1}^n x_i s_i - n \overline{x} \overline{s}) \sum\limits_{i=1}^n s_i
    \end{matrix} \right) \\
    & \left( \sum\limits_{i=1}^n x_i y_i - n \overline{x} \overline{y}, \sum\limits_{i=1}^n x_i s_i - n \overline{x} \overline{s} \right)^\top \\
\end{split}
\]

\[
\begin{split}
    & = a \cdot b \cdot n \\
    &
    \left(
    \begin{matrix}
    \underbrace{\sum\limits_{i=1}^n x_i y_i \sum\limits_{i=1}^n (s_i - \overline{s})^2 - n \overline{x} \overline{y} \sum\limits_{i=1}^n (s_i - \overline{s})^2 - \sum\limits_{i=1}^n x_i s_i \sum\limits_{i=1}^n \sqrt{s_i} (s_i - \overline{s}) - n \overline{x} \overline{s} \sum\limits_{i=1}^n \sqrt{s_i} (s_i - \overline{s}),}_{:=k} \\
    \underbrace{- \sum\limits_{i=1}^n x_i y_i \sum\limits_{i=1}^n \sqrt{s_i} (s_i - \overline{s}) + n \overline{x} \overline{y} \sum\limits_{i=1}^n \sqrt{s_i} (s_i - \overline{s}) + \sum\limits_{i=1}^n x_i s_i \sum\limits_{i=1}^n s_i - n \overline{x} \overline{s} \sum\limits_{i=1}^n s_i}_{:=m}
    \end{matrix}
    \right) \\
    & \left( \sum\limits_{i=1}^n x_i y_i - n \overline{x} \overline{y}, \sum\limits_{i=1}^n x_i s_i - n \overline{x} \overline{s} \right)^\top \\
    & = a \cdot b \cdot n \cdot (k, m) \cdot
    \left( \sum\limits_{i=1}^n x_i y_i - n \overline{x} \overline{y}, \sum\limits_{i=1}^n x_i s_i - n \overline{x} \overline{s} \right)^\top \\
    & = a \cdot b \cdot n \cdot
    \left( k (\sum\limits_{i=1}^n x_i y_i - n \overline{x} \overline{y}) + m (\sum\limits_{i=1}^n x_i s_i - n \overline{x} \overline{s}) \right) \\
\end{split}
\]

\subsubsection*{DistTree}
Since score functions are used in the test statistic for DistTree, we will define them first \citep[cf.][]{fahrmeir2016statistik}:
$$
s(\boldsymbol{\hat{\mu}}, y_i) = \frac{y_i - \hat{\mu}}{\hat{\sigma}^2};  \quad 
s(\boldsymbol{\hat{\sigma}}, y_i) = -\frac{1}{\hat{\sigma}} + \frac{(y_i - \hat{\mu})^2}{\hat{\sigma}^3}.
$$

From the maximum likelihood estimation it follows that 
$$
\hat{\mu} = \overline{y};  \quad 
\hat{\sigma} = \sqrt{\frac{1}{n} \sum\limits_{i=1}^n (y_i - \overline{y})^2} = \sqrt{\overline{s}}.
$$

Therefore, we can express the score functions as follows:
$$
s(\boldsymbol{\hat{\mu}}, y_i) = \frac{y_i - \overline{y}}{\overline{s}} = \frac{\sqrt{s_i}}{\overline{s}};  \quad 
s(\boldsymbol{\hat{\sigma}}, y_i) = -\frac{1}{\sqrt{\overline{s}}} + \frac{\overbrace{(y_i - \overline{y})^2}^{=s_i}}{\overline{s} \sqrt{\overline{s}}} = \frac{s_i - \overline{s}}{\overline{s} \sqrt{\overline{s}}}.
$$

In DistTree the statistic

\[
 \begin{split}
    t & = vec \left( \sum\limits_{i=1}^n g(x_i) s(\boldsymbol{\hat{\theta}}, y_i)  \right) = \sum\limits_{i=1}^n x_i (s(\boldsymbol{\hat{\mu}}, y_i), s(\boldsymbol{\hat{\sigma}}, y_i)) \\
      & = \left( \sum\limits_{i=1}^n x_i s(\boldsymbol{\hat{\mu}}, y_i), \sum\limits_{i=1}^n x_i s(\boldsymbol{\hat{\sigma}}, y_i) \right) = \left( \sum\limits_{i=1}^n x_i \frac{\sqrt{s_i}}{\overline{s}}, \sum\limits_{i=1}^n x_i \frac{s_i - \overline{s}}{\overline{s} \sqrt{\overline{s}}} \right)
 \end{split}
\]

has expectation

\[
 \begin{split}
    \mu & = vec \left( \sum\limits_{i=1}^n g(x_i) \frac{1}{n} \sum\limits_{i=1}^n (s(\boldsymbol{\hat{\mu}}, y_i), s(\boldsymbol{\hat{\sigma}}, y_i))  \right) = \left( \frac{1}{n} \sum\limits_{i=1}^n x_i \sum\limits_{i=1}^n\frac{\sqrt{s_i}}{\overline{s}}, \frac{1}{n} \sum\limits_{i=1}^n x_i \sum\limits_{i=1}^n \frac{s_i - \overline{s}}{\overline{s} \sqrt{\overline{s}}} \right) \\
      & = \left( \overline{x} \sum\limits_{i=1}^n \frac{\sqrt{s_i}}{\overline{s}}, \overline{x} \sum\limits_{i=1}^n \frac{s_i - \overline{s}}{\overline{s} \sqrt{\overline{s}}} \right).
 \end{split}
\]

The covariance $\Sigma$ is defined similarly to CTreeTrafo (see Equation (\ref{eq3})), where $V$ is defined as follows:

\[
\begin{split}
    V & = \frac{1}{n} \sum\limits_{i=1}^n \left( \left(\frac{y_i - \overline{y}}{\overline{s}}, \frac{s_i - \overline{s}}{\overline{s} \sqrt{\overline{s}}}\right) - \frac{1}{n} \sum\limits_{i=1}^n \left(\frac{y_i - \overline{y}}{\overline{s}}, \frac{s_i - \overline{s}}{\overline{s} \sqrt{\overline{s}}}\right) \right) \left( \left(\frac{y_i - \overline{y}}{\overline{s}}, \frac{s_i - \overline{s}}{\overline{s} \sqrt{\overline{s}}}\right) - \frac{1}{n} \sum\limits_{i=1}^n \left(\frac{y_i - \overline{y}}{\overline{s}}, \frac{s_i - \overline{s}}{\overline{s} \sqrt{\overline{s}}}\right) \right)^\top \\
     & = \frac{1}{n} \sum\limits_{i=1}^n \left( \frac{1}{\overline{s}} \left(y_i - \overline{y} - \frac{1}{n} \sum\limits_{i=1}^n y_i + \frac{1}{n} \sum\limits_{i=1}^n \overline{y} \right), \frac{1}{\overline{s} \sqrt{\overline{s}}} \left(s_i - \overline{s} - \frac{1}{n} \sum\limits_{i=1}^n s_i + \frac{1}{n} \sum\limits_{i=1}^n \overline{s} \right) \right) \\
     & \left( \frac{1}{\overline{s}} \left(y_i - \overline{y} - \frac{1}{n} \sum\limits_{i=1}^n y_i + \frac{1}{n} \sum\limits_{i=1}^n \overline{y} \right), \frac{1}{\overline{s} \sqrt{\overline{s}}} \left(s_i - \overline{s} - \frac{1}{n} \sum\limits_{i=1}^n s_i + \frac{1}{n} \sum\limits_{i=1}^n \overline{s} \right) \right)^\top \\
     & = \frac{1}{n} \sum\limits_{i=1}^n \left( \frac{1}{\overline{s}} \left(y_i - \overline{y} - \overline{y} + \overline{y} \right), \frac{1}{\overline{s} \sqrt{\overline{s}}} \left(s_i - \overline{s} - \overline{s} + \overline{s} \right) \right) \left( \frac{1}{\overline{s}} \left(y_i - \overline{y} - \overline{y} + \overline{y} \right), \frac{1}{\overline{s} \sqrt{\overline{s}}} \left(s_i - \overline{s} - \overline{s} + \overline{s} \right) \right)^\top \\
     & = \frac{1}{n} \sum\limits_{i=1}^n \left( \frac{1}{\overline{s}} \left(y_i - \overline{y} \right), \frac{1}{\overline{s} \sqrt{\overline{s}}} \left(s_i - \overline{s} \right) \right) \left( \frac{1}{\overline{s}} \left(y_i - \overline{y} \right), \frac{1}{\overline{s} \sqrt{\overline{s}}} \left(s_i - \overline{s} \right) \right)^\top \\
      & = \frac{1}{n}
      \left(
      \begin{matrix}
        \frac{1}{\overline{s}^2} \sum\limits_{i=1}^n (y_i - \overline{y})^2 & \frac{1}{\overline{s}^2 \sqrt{\overline{s}}} \sum\limits_{i=1}^n (y_i - \overline{y})(s_i - \overline{s}) \\
      \frac{1}{\overline{s}^2 \sqrt{\overline{s}}} \sum\limits_{i=1}^n (y_i - \overline{y})(s_i - \overline{s}) & \frac{1}{\overline{s}^2 \overline{s}} \sum\limits_{i=1}^n (s_i - \overline{s})^2
      \end{matrix} 
      \right) \\
      & = \frac{1}{n} \frac{1}{\overline{s}^2}
      \left(
      \begin{matrix}
        \sum\limits_{i=1}^n s_i & \frac{1}{\sqrt{\overline{s}}} \sum\limits_{i=1}^n \sqrt{s_i}(s_i - \overline{s}) \\
      \frac{1}{\sqrt{\overline{s}}} \sum\limits_{i=1}^n \sqrt{s_i}(s_i - \overline{s}) & \frac{1}{\overline{s}} \sum\limits_{i=1}^n (s_i - \overline{s})^2
      \end{matrix}
      \right)
 \end{split}
\]

\newpage
Combined we get

\[
\begin{split}
    c_{quad}(t, \mu, \Sigma) 
    & = \left( t - \mu \right) \Sigma^{-1} \left( t - \mu \right)^\top \\
    & = \left( \frac{1}{\overline{s}} (\sum\limits_{i=1}^n x_i y_i - n \overline{x} \overline{y}), \frac{1}{\overline{s} \sqrt{\overline{s}}} (\sum\limits_{i=1}^n x_i s_i - n \overline{x} \overline{s}) \right) \Sigma^{-1} \\
    & \left( \frac{1}{\overline{s}} (\sum\limits_{i=1}^n x_i y_i - n \overline{x} \overline{y}), \frac{1}{\overline{s} \sqrt{\overline{s}}} (\sum\limits_{i=1}^n x_i s_i - n \overline{x} \overline{s}) \right)^\top \\
    & = \left( \frac{1}{\overline{s}} (\sum\limits_{i=1}^n x_i y_i - n \overline{x} \overline{y}), \frac{1}{\overline{s} \sqrt{\overline{s}}} (\sum\limits_{i=1}^n x_i s_i - n \overline{x} \overline{s}) \right) \left( \frac{n}{n-1} V \left( \sum\limits_{i=1}^n - n\overline{x}^2 \right) \right)^{-1} \\
    & \left( \frac{1}{\overline{s}} (\sum\limits_{i=1}^n x_i y_i - n \overline{x} \overline{y}), \frac{1}{\overline{s} \sqrt{\overline{s}}} (\sum\limits_{i=1}^n x_i s_i - n \overline{x} \overline{s}) \right)^\top \\
    & = \left( \frac{1}{\overline{s}} (\sum\limits_{i=1}^n x_i y_i - n \overline{x} \overline{y}), \frac{1}{\overline{s} \sqrt{\overline{s}}} (\sum\limits_{i=1}^n x_i s_i - n \overline{x} \overline{s}) \right) \underbrace{\left( \frac{n}{n-1} \left( \sum\limits_{i=1}^n - n\overline{x}^2 \right) \right)^{-1}}_{:=a} V^{-1} \\
    & \left( \frac{1}{\overline{s}} (\sum\limits_{i=1}^n x_i y_i - n \overline{x} \overline{y}), \frac{1}{\overline{s} \sqrt{\overline{s}}} (\sum\limits_{i=1}^n x_i s_i - n \overline{x} \overline{s}) \right)^\top \\
    & = \left( \frac{1}{\overline{s}} (\sum\limits_{i=1}^n x_i y_i - n \overline{x} \overline{y}), \frac{1}{\overline{s} \sqrt{\overline{s}}} (\sum\limits_{i=1}^n x_i s_i - n \overline{x} \overline{s}) \right) a \\
    & \underbrace{\frac{1}{\sum\limits_{i=1}^n s_i \frac{1}{\overline{s}} \sum\limits_{i=1}^n (s_i - \overline{s})^2 - (\frac{1}{\sqrt{\overline{s}}} \sum\limits_{i=1}^n \sqrt{s_i}(s_i - \overline{s}))^2}}_{(\star_1)} n \cdot \overline{s}^2 \\
    &
    \left(
    \begin{matrix}
       \frac{1}{\overline{s}} \sum\limits_{i=1}^n (s_i - \overline{s})^2 & \frac{1}{\sqrt{\overline{s}}} -\sum\limits_{i=1}^n \sqrt{s_i} (s_i - \overline{s}) \\
        \frac{1}{\sqrt{\overline{s}}} -\sum\limits_{i=1}^n \sqrt{s_i} (s_i - \overline{s}) & \sum\limits_{i=1}^n s_i
    \end{matrix}
    \right)
     \left( \frac{1}{\overline{s}} (\sum\limits_{i=1}^n x_i y_i - n \overline{x} \overline{y}), \frac{1}{\overline{s} \sqrt{\overline{s}}} (\sum\limits_{i=1}^n x_i s_i - n \overline{x} \overline{s}) \right)^\top \\
     & = a \cdot b \cdot n \cdot \overline{s}^3
     \left( \frac{1}{\overline{s}} (\sum\limits_{i=1}^n x_i y_i - n \overline{x} \overline{y}), \frac{1}{\overline{s} \sqrt{\overline{s}}} (\sum\limits_{i=1}^n x_i s_i - n \overline{x} \overline{s}) \right) \\
     & \left(
     \begin{matrix}
       \frac{1}{\overline{s}} \sum\limits_{i=1}^n (s_i - \overline{s})^2 & \frac{1}{\sqrt{\overline{s}}} -\sum\limits_{i=1}^n \sqrt{s_i} (s_i - \overline{s}) \\
       \frac{1}{\sqrt{\overline{s}}} -\sum\limits_{i=1}^n \sqrt{s_i} (s_i - \overline{s}) & \sum\limits_{i=1}^n s_i
    \end{matrix}
    \right)
    \left( \frac{1}{\overline{s}} (\sum\limits_{i=1}^n x_i y_i - n \overline{x} \overline{y}), \frac{1}{\overline{s} \sqrt{\overline{s}}} (\sum\limits_{i=1}^n x_i s_i - n \overline{x} \overline{s}) \right)^\top \\
\end{split}
\]

\[
\begin{split}
    & = a \cdot b \cdot n \cdot \overline{s}^3
    \underbrace{\left( \begin{matrix}
    \frac{1}{\overline{s}} (\sum\limits_{i=1}^n x_i y_i - n \overline{x} \overline{y}) \frac{1}{\overline{s}} \sum\limits_{i=1}^n (s_i - \overline{s})^2 - \frac{1}{\overline{s}\sqrt{\overline{s}}} (\sum\limits_{i=1}^n x_i s_i - n \overline{x} \overline{s}) \frac{1}{\sqrt{\overline{s}}} \sum\limits_{i=1}^n \sqrt{s_i} (s_i - \overline{s}), \\
    -  \frac{1}{\overline{s}} (\sum\limits_{i=1}^n x_i y_i - n \overline{x} \overline{y}) \frac{1}{\sqrt{\overline{s}}} \sum\limits_{i=1}^n \sqrt{s_i} (s_i - \overline{s}) + \frac{1}{\overline{s}\sqrt{\overline{s}}} (\sum\limits_{i=1}^n x_i s_i - n \overline{x} \overline{s}) \sum\limits_{i=1}^n s_i
    \end{matrix} \right)}_{(\star_2)} \\
    & \left( \frac{1}{\overline{s}} (\sum\limits_{i=1}^n x_i y_i - n \overline{x} \overline{y}), \frac{1}{\overline{s} \sqrt{\overline{s}}} (\sum\limits_{i=1}^n x_i s_i - n \overline{x} \overline{s}) \right)^\top \\
    & = a \cdot b \cdot n \cdot \overline{s}^3 \cdot \left( \frac{1}{\overline{s}^2} k, \frac{1}{\overline{s}\sqrt{\overline{s}}} m \right) \cdot
    \left( \frac{1}{\overline{s}} (\sum\limits_{i=1}^n x_i y_i - n \overline{x} \overline{y}), \frac{1}{\overline{s} \sqrt{\overline{s}}} (\sum\limits_{i=1}^n x_i s_i - n \overline{x} \overline{s}) \right)^\top \\
    & = a \cdot b \cdot n \cdot \overline{s}^3 \cdot
    \left( \frac{1}{\overline{s}^2} k \frac{1}{\overline{s}} (\sum\limits_{i=1}^n x_i y_i - n \overline{x} \overline{y}) + \frac{1}{\overline{s}\sqrt{\overline{s}}} m \frac{1}{\overline{s}\sqrt{\overline{s}}} (\sum\limits_{i=1}^n x_i s_i - n \overline{x} \overline{s}) \right) \\
    & = a \cdot b \cdot n \cdot \overline{s}^3 \cdot
    \left( \frac{1}{\overline{s}^3} k (\sum\limits_{i=1}^n x_i y_i - n \overline{x} \overline{y}) + \frac{1}{\overline{s}^3} m (\sum\limits_{i=1}^n x_i s_i - n \overline{x} \overline{s}) \right) \\
    & = a \cdot b \cdot n \cdot \overline{s}^3 \cdot \frac{1}{\overline{s}^3} \cdot
    \left( k (\sum\limits_{i=1}^n x_i y_i - n \overline{x} \overline{y}) + m (\sum\limits_{i=1}^n x_i s_i - n \overline{x} \overline{s}) \right) \\
    & = a \cdot b \cdot n \cdot
    \left( k (\sum\limits_{i=1}^n x_i y_i - n \overline{x} \overline{y}) + m (\sum\limits_{i=1}^n x_i s_i - n \overline{x} \overline{s}) \right) \\
\end{split}
\]
\\

In the following we resolve the individual components ($\star_1$, $\star_2$) of $c_{quad}$.

\[
\begin{split}
    \star_1 & = \frac{1}{\frac{1}{\overline{s}} \sum\limits_{i=1}^n s_i \sum\limits_{i=1}^n (s_i - \overline{s})^2 - \frac{1}{\overline{s}} ( \sum\limits_{i=1}^n \sqrt{s_i}(s_i - \overline{s}))^2} \\
    & = \frac{1}{\frac{1}{\overline{s}}} \cdot \underbrace{\frac{1}{ \sum\limits_{i=1}^n s_i \sum\limits_{i=1}^n (s_i - \overline{s})^2 - ( \sum\limits_{i=1}^n \sqrt{s_i}(s_i - \overline{s}))^2}}_{:=b} \\
    & = \overline{s} \cdot b \\
\end{split}
\]

\[
\begin{split}
    \star_2 & = \left( \begin{matrix}
    \frac{1}{\overline{s}} \sum\limits_{i=1}^n x_i y_i \frac{1}{\overline{s}} \sum\limits_{i=1}^n (s_i - \overline{s})^2 - 
    \frac{1}{\overline{s}} n \overline{x} \overline{y} \frac{1}{\overline{s}} \sum\limits_{i=1}^n (s_i - \overline{s})^2 - \frac{1}{\overline{s}\sqrt{\overline{s}}} \sum\limits_{i=1}^n x_i s_i \frac{1}{\sqrt{\overline{s}}} \sum\limits_{i=1}^n \sqrt{s_i} (s_i - \overline{s}) \\
    + \frac{1}{\overline{s}\sqrt{\overline{s}}} n \overline{x} \overline{s} \frac{1}{\sqrt{\overline{s}}} \sum\limits_{i=1}^n \sqrt{s_i} (s_i - \overline{s}), \\
    - \frac{1}{\overline{s}} \sum\limits_{i=1}^n x_i y_i \frac{1}{\sqrt{\overline{s}}} \sum\limits_{i=1}^n \sqrt{s_i} (s_i - \overline{s}) + \frac{1}{\overline{s}} n \overline{x} \overline{y} \frac{1}{\sqrt{\overline{s}}} \sum\limits_{i=1}^n \sqrt{s_i} (s_i - \overline{s}) + \frac{1}{\overline{s}\sqrt{\overline{s}}} \sum\limits_{i=1}^n x_i s_i \sum\limits_{i=1}^n s_i - \frac{1}{\overline{s}\sqrt{\overline{s}}} n \overline{x} \overline{s} \sum\limits_{i=1}^n s_i
    \end{matrix} \right)\\
     & = \left( \begin{matrix}
    \frac{1}{\overline{s}^2} \sum\limits_{i=1}^n x_i y_i \sum\limits_{i=1}^n (s_i - \overline{s})^2 - 
    \frac{1}{\overline{s}^2} n \overline{x} \overline{y} \sum\limits_{i=1}^n (s_i - \overline{s})^2 - \frac{1}{\overline{s}^2} \sum\limits_{i=1}^n x_i s_i \sum\limits_{i=1}^n \sqrt{s_i} (s_i - \overline{s}) + \frac{1}{\overline{s}^2} n \overline{x} \overline{s} \sum\limits_{i=1}^n \sqrt{s_i} (s_i - \overline{s}), \\
    - \frac{1}{\overline{s}\sqrt{\overline{s}}} \sum\limits_{i=1}^n x_i y_i \sum\limits_{i=1}^n \sqrt{s_i} (s_i - \overline{s}) + \frac{1}{\overline{s}\sqrt{\overline{s}}} n \overline{x} \overline{y} \sum\limits_{i=1}^n \sqrt{s_i} (s_i - \overline{s}) + \frac{1}{\overline{s}\sqrt{\overline{s}}} \sum\limits_{i=1}^n x_i s_i \sum\limits_{i=1}^n s_i - \frac{1}{\overline{s}\sqrt{\overline{s}}} n \overline{x} \overline{s} \sum\limits_{i=1}^n s_i
    \end{matrix} \right)\\
    & = \left( \begin{matrix}
    \frac{1}{\overline{s}^2} \left( \underbrace{\sum\limits_{i=1}^n x_i y_i \sum\limits_{i=1}^n (s_i - \overline{s})^2 - 
    n \overline{x} \overline{y} \sum\limits_{i=1}^n (s_i - \overline{s})^2 - \sum\limits_{i=1}^n x_i s_i \sum\limits_{i=1}^n \sqrt{s_i} (s_i - \overline{s}) + n \overline{x} \overline{s} \sum\limits_{i=1}^n \sqrt{s_i} (s_i - \overline{s})}_{:=k} \right), \\
    \frac{1}{\overline{s}\sqrt{\overline{s}}} \left( \underbrace{- \sum\limits_{i=1}^n x_i y_i \sum\limits_{i=1}^n \sqrt{s_i} (s_i - \overline{s}) + n \overline{x} \overline{y} \sum\limits_{i=1}^n \sqrt{s_i} (s_i - \overline{s}) + \sum\limits_{i=1}^n x_i s_i \sum\limits_{i=1}^n s_i - n \overline{x} \overline{s} \sum\limits_{i=1}^n s_i}_{:=m} \right)
    \end{matrix} \right)\\
    & = \left( \frac{1}{\overline{s}^2} k, \frac{1}{\overline{s}\sqrt{\overline{s}}} m \right) \\
\end{split}
\]

Substituting the above components into the statistics $c_{quad}$ of CTreeTrafo and DistTree, we find that both statistics are

$$
 a \cdot b \cdot n \cdot
    \left( k (\sum\limits_{i=1}^n x_i y_i - n \overline{x} \overline{y}) + m (\sum\limits_{i=1}^n x_i s_i - n \overline{x} \overline{s}) \right). 
$$

\end{document}